\documentclass[pra,aps,twocolumn,nopacs,superscriptaddress,nofootinbib]{revtex4}
\usepackage{graphicx}
\usepackage{dcolumn}
\usepackage{bm}
\usepackage{amsmath}
\usepackage{epsfig}
\usepackage{color}

\def\ii{{\rm i}}  \def\ee{{\rm e}}    
\def\Rb{{\bf R}}    \def\Qb{{\bf Q}}  
      \def\rb{{\bf r}}

\def\Ef{{E_{\text{F}}}}           
  				\def\kp{{k_{\parallel}}}
	  				\def\kpb{{\bf k_{\parallel}}}	   
\def\vf{{v_{\text{F}}}}   \def\kf{{k_{\text{F}}}}   \def\EF{{E_{\text{F}}}}
  \def\Vb{{V_{\text{b}}}}
\def\phiwell{{\varphi^\perp}}

\begin{document}

\title{Plasmon Generation through Electron Tunneling in Graphene}

\author{Sandra~de~Vega}
\affiliation{ICFO-Institut de Ciencies Fotoniques, The Barcelona Institute of Science and Technology, 08860 Castelldefels (Barcelona), Spain}
\author{F.~Javier~Garc\'{\i}a~de~Abajo}
\affiliation{ICFO-Institut de Ciencies Fotoniques, The Barcelona Institute of Science and Technology, 08860 Castelldefels (Barcelona), Spain}
\affiliation{ICREA-Instituci\'o Catalana de Recerca i Estudis Avan\c{c}ats, Passeig Llu\'{\i}s Companys 23, 08010 Barcelona, Spain}
\email{javier.garciadeabajo@nanophotonics.es}

\begin{abstract}
{\bf The short wavelength of graphene plasmons relative to the light wavelength makes them attractive for applications in optoelectronics and sensing. However, this property limits their coupling to external light and our ability to create and detect them. More efficient ways of generating plasmons are therefore desirable. Here we demonstrate through realistic theoretical simulations that graphene plasmons can be efficiently excited {\it via} electron tunneling in a sandwich structure formed by two graphene monolayers  separated by a few atomic layers of hBN. We predict plasmon generation rates of $\sim10^{12}-10^{14}/$s over an area of the squared plasmon wavelength for realistic values of the spacing and bias voltage, while the yield (plasmons per tunneled electron) has unity order. Our results support electrical excitation of graphene plasmons in tunneling devices as a viable mechanism for the development of optics-free ultrathin plasmonic devices.}
\end{abstract}
\date{\today}
\maketitle



\section{Introduction}

Plasmons offer the means to concentrate optical fields down to nanometer-sized regions and enhance the intensity of externally incident light by several orders of magnitude \cite{LSB03,paper156}. These properties have been extensively investigated to develop applications in areas as diverse as optical sensing \cite{LNL1983,AHL08,ZBH14,RJL17}, medical diagnosis and treatment \cite{NHH04,JEE07,LSH11}, nonlinear optics \cite{PN08,PDN09}, catalysis \cite{SZY12,WRN13,C14,PBS15}, and photovoltaics \cite{CP08,AP10}. In this context, plasmons in high-quality graphene offer the advantages of being electrically \cite{FAB11,FRA12,paper196,paper212,BJS13,BSS14}, magnetically \cite{YLL12,KRR14}, and optically \cite{paper235,NWG16} tunable, while exhibiting short wavelengths and long lifetimes \cite{WLG15}. Graphene plasmons are highly customizable through lateral patterning, stacking of 2D-crystal atomic layers, and coupling to the dielectric environment \cite{paper283}, enabling applications such as light modulation \cite{YLZ12,BJS13}, control of thermal emission \cite{FCS10,IJJ12,paper245,BSJ15}, spectral photometry \cite{paper275,LGW16}, and optical sensing \cite{LYF14,paper255,paper256,FAL16}. While these phenomena and applications have been mainly explored at mid-infrared and lower frequencies, prospects for their extension to the visible and near-infrared spectral regions look promising \cite{paper235}. Actually, there is no fundamental reason that limits the existence of graphene plasmons at such high frequencies, although in practice they require large doping ({\it e.g.}, an attainable $\EF\sim1\,$eV Fermi energy \cite{CPB11}) and small lateral size $<10\,$nm. Although the latter is challenging using top-down lithography, exciting possibilities are opened by bottom-up approaches to the synthesis of nanographenes with only a few nanometers in lateral size \cite{YXL13,M14}, while recent experiments on self-assembled graphene nanodisks already reveal plasmons close to the near-infrared \cite{WLA16}. In fact, plasmon-like resonances have been measured in small aromatic hydrocarbons \cite{paper215,paper260}, which can be considered as molecular versions of graphene.

The strong confinement of graphene plasmons is clearly indicated by the ratio of plasmon-to-light wavelengths for a self-standing carbon monolayer \cite{paper235}, $\lambda_{\rm p}/\lambda=2\alpha(\EF/\hbar\omega)\ll1$, where $\alpha\approx1/137$ is the fine structure constant and $\omega$ is the light frequency. This relation implies that graphene plasmons can be described in the quasistatic limit. Unfortunately, it also means that their in/out-coupling to propagating light is weak. Let us emphasize that far-field coupling in homogeneous graphene is forbidden because plasmons are evanescent surface waves with energy and momentum outside the light cone; however, we refer here to the limited ability of graphene nanostructures to produce in/out plasmon coupling, which is neatly illustrated by their optical extinction and elastic-scattering cross-sections. Indeed, a simple extension of previous analytical theory \cite{paper235} based on the Drude conductivity indicates that the radiative decay rate of a low-order plasmon in a graphene structure ({\it e.g.}, a disk) is $\kappa_{\rm r}\sim c A \EF/\lambda^3 E_{\rm p}$, which for a plasmon energy $E_{\rm p}$ comparable with the Fermi energy $\EF$, a graphene area $A\sim\lambda_{\rm p}^2\sim10 ^{-4}\lambda^2$, and a characteristic photon wavelength of $2\,\mu$m, leads to $\kappa_{\rm r}\sim0.1/$ns. This is orders of magnitude smaller than the total plasmon decay rate $\kappa$. Now, even for an unrealistically optimistic value $\kappa=1/$ps, the on-resonance plasmon-driven extinction ($\sigma^{\rm ext}=(3\lambda^2/2\pi)\kappa_{\rm r}/\kappa\sim10^{-4}\lambda^2$) and elastic-scattering ($\sigma^{\rm scat}=(\kappa_{\rm r}/\kappa)\sigma^{\rm ext}\sim10^{-8}\lambda^2$) cross-sections are small, thus rendering plasmon coupling to propagating light as a mere retardation correction. In order to circumvent this limitation, many experimental studies in this field have relied on near-field nanoscopy \cite{FNJ17}, which is based on the use of sharp tips to enhance this coupling \cite{FRA12,paper196,NWG16}. However, more compact devices for the generation and detection of graphene plasmons are needed in order to enable the design and development of applications in integrated architectures.

A promising approach consists in exploiting inelastic electron transitions to excite and detect plasmons. Understandably, focused beams in electron microscopes have been the probe of choice to excite and map plasmons with nanometer spatial resolution {\it via} the recorded electron energy-loss and cathodoluminescence signals \cite{paper149}. Electron tunneling has been considered for a long time as a mechanism for the excitation of plasmons \cite{T1969,TB1969,D1969,DRS1969,EN1971,JMA1990,BGJ91,RHB13,OCH15}, while electron injection by tunneling from conducting tips into metallic structures has been also demonstrated to produce efficient plasmon excitation \cite{SSB10,BBN11,GKL14,KKP15,UKB17}. More dedicated designs have incorporated emitting devices in which the generated light directly couples to the plasmon near-field \cite{KHD08,WVB10}. A recent theoretical study has shown that plasmons can boost tunneling across an insulator separating two graphene layers \cite{EBS17}, with potential use as a plasmon-gain device \cite{SDR16}, while a similar structure has been experimentally used for the electrical generation of THz radiation \cite{YTW16}. Additionally, evidence of radiative gap-plasmon decay has been experimentally obtained associated with hot electron tunneling under external illumination \cite{WBZ15}. 

In a related context, efforts to realize electrical detection of plasmons have relied on near-field coupling to structures consisting of organic diodes \cite{DAK06}, superconductors \cite{HDK09_2}, nanowire semiconductors \cite{NVD09,FKC09}, Schottky barriers \cite{DDV10,GDK11,KSN11}, and 2D semiconductors \cite{GCB15}. Thermoelectric detection has been recently demonstrated in graphene \cite{LGW16}, while a nanoscale detector has been proposed based on changes produced in the conductivity of graphene nanojunctions \cite{paper275}. These technologies should benefit from advances in the design, nanofabrication, and theoretical modeling of transistors made from 2D material sandwiches \cite{BGG13,JOB16,KO17}.

Here, we theoretically investigate a simple structure in which two closely spaced graphene sheets serve both as gates to produce electron tunneling and as plasmon-supporting elements. We predict a generation efficiency that approaches one plasmon per tunneled electron. Considering attainable doping conditions and bias voltages applied to the graphene layers, we find a generation rate of $\sim10^{12}-10^{14}$ plasmons per second over an area of a squared plasmon wavelength for a 1\,nm inter-graphene spacing filled with hexagonal boron nitride (hBN). This type of structure provides a key element for future optics-free integrated devices and could also be operated in reversed mode to detect plasmons by decay into tunneled electrons.

\begin{figure}
\begin{center}
\includegraphics[width=0.5\textwidth]{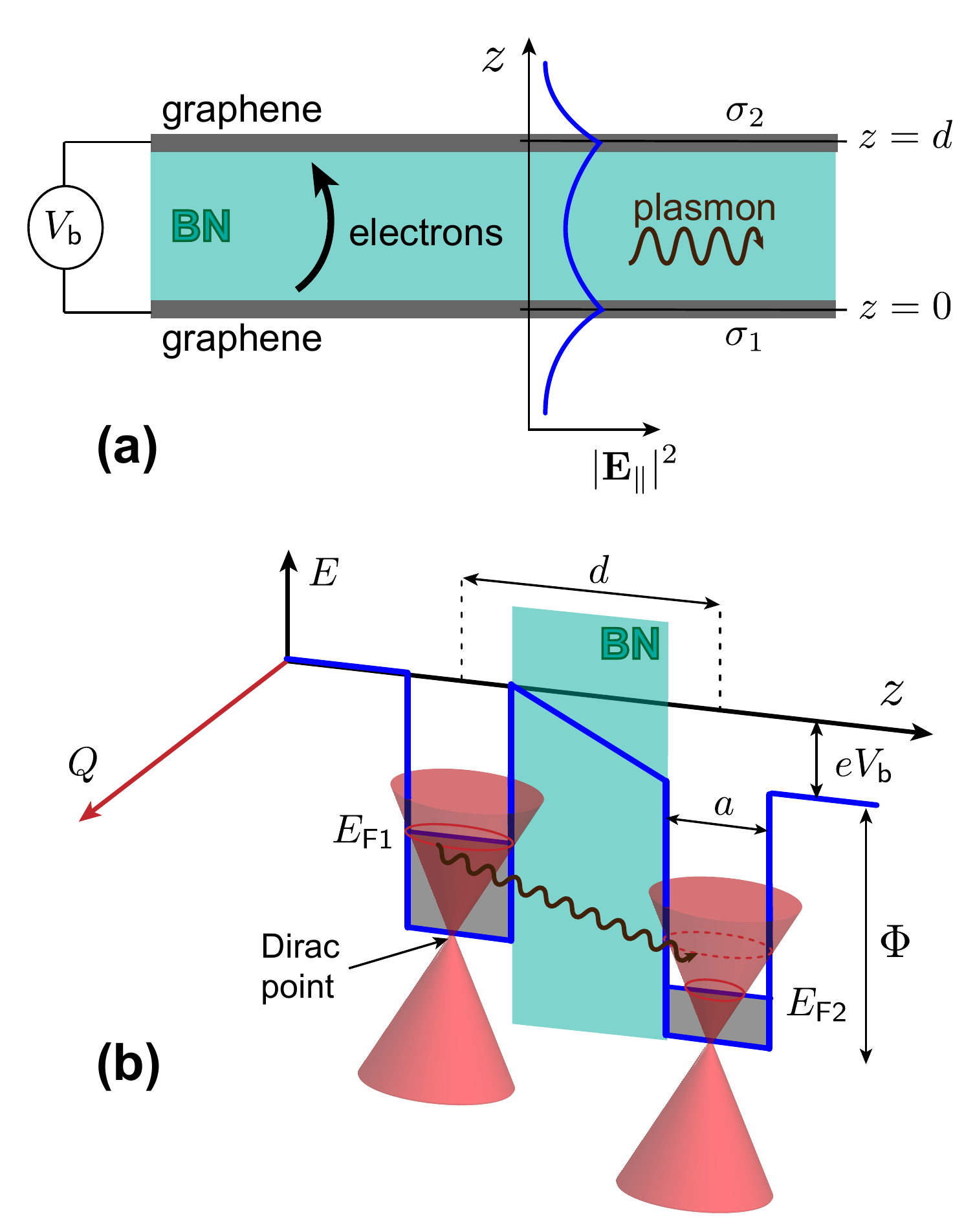}
\caption{{\bf Electron Tunneling in Double-Layer Graphene}. {\bf (a)} We consider a graphene-hBN-graphene sandwich structure consisting of two crystallographically-aligned monolayer graphene sheets separated by a hBN film of thickness $d$. Electron tunneling takes place upon application of a bias voltage $\Vb$. The graphene layers are described through their wave-vector- and frequency-dependent conductivities  $\sigma_1(\kp,\omega)$ and $\sigma_2(\kp,\omega)$. The electric-field intensity associated with the symmetric optical plasmon sustained by this structure is plotted schematically. {\bf (b)} We show the electronic bands of the graphene layers and their relative energy positions due to the bias. The Fermi energies $\Ef_1$ and $\Ef_2$ and the work function $\Phi$ are indicated. The electron wave functions have a dependence along the film normal direction modeled through two quantum wells (blue profile), while Dirac fermions describe their dependence on parallel directions (red cones and parallel electron wave vector $Q$). Electron tunneling (wiggly arrow) is limited by parallel momentum conservation.} 
\label{Fig1}
\end{center}
\end{figure}

\section{Plasmon Generation through Electron Tunneling}

The structure under consideration (Fig.\ \ref{Fig1}a) consists of two graphene layers separated by a hBN film and placed at different Fermi energies $\EF_1$ and $\EF_2$. Upon application of a bias voltage $\Vb$ between the two carbon layers, electrons can tunnel from one to the other, assisted by the excitation of a propagating plasmon (inelastic tunneling). We schematically depict the profile of a symmetric plasmon superimposed on the scheme ({\it i.e.}, an excitation to which the tunneling electrons couple with high efficiency, see below).

Elastic tunneling ({\it i.e.}, direct tunneling of electrons from one layer to the other) is forbidden due to the vertical displacement of their respective Dirac cones associated with the bias, which introduces a mismatch of parallel wave vectors for all electron energies (see Fig.\ \ref{Fig1}b). In contrast, the creation of a plasmon (wiggly arrow) can break this mismatch and enable inelastic tunneling of the electron. Nonetheless, this process involves only a specific energy in the graphene electron bands for each inelastic frequency $\omega$, which is controlled through the applied bias voltage and the Fermi energies of the two carbon layers. The theoretical investigation of plasmon excitation during electron tunneling has been pioneered in the context of junctions between semiconductors using the self-energy \cite{DRS1969}. A recent theoretical study\cite{EBS17} discusses in great detail plasmon generation associated with electron tunneling using a second-quantization formalism. We follow instead a more direct approach based upon methods borrowed from the analysis of electron energy-loss spectroscopy \cite{paper149}.

The elements involved in the theoretical description of inelastic electron tunneling are {\it (i)} the conduction electron wave functions; {\it (ii)} the screened Coulomb potential $W$ that mediates the interaction between those electrons; and {\it (iii)} the evaluation of the transition probability through a well-established linear-response formalism \cite{paper149}. We present a a self-contained derivation in the Appendix, under the assumption of crystallographically aligned graphene sheets ({\it i.e.}, their Dirac cones are on top of each other in 2D momentum space). In particular, electrons are assumed to have their wave functions factorized as the product of components parallel (Dirac fermions \cite{CGP09}) and perpendicular to the surface. The latter accounts for the transversal confinement to the graphene layers, which we describe through a simple potential-well model (blue profile in Fig.\ \ref{Fig1}b), with parameters fitted to match the work function and electron spill out of $p_z$ orbitals in the material (see Appendix). Actually, spill out is a crucial element because it determines the overlap of electronic wave functions between the two graphene layers. For simplicity, we neglect the repulsive electron potential in the hBN region, which should produce a correction to this overlap. The inelastic tunneling current density is then expressed as an integral over parallel electron wave vectors (see Eq.\ (\ref{Gkw})), which involves the Fermi-Dirac occupation distributions of the graphene bands in the two layers (we assume a temperature of $300\,$K). The screened Coulomb interaction enters this integral and incorporates the optical responses of the graphene layers ({\it i.e.}, the corresponding conductivities) and the hBN film (see Appendix). We use the random-phase approximation \cite{W1947,WSS06,HD07} (RPA) to describe the frequency- and wave-vector-dependence of the graphene conductivities, thus incorporating nonlocal effects that are important because both interlayer dielectric coupling and electron tunneling involve large values of the transferred parallel momentum for the short separations under consideration. The conductivities depend on the Fermi energies of the graphene layers, as well as on the electron lifetime, which we set to a conservative value $\tau=66\,$fs. We remark that our results for the integral of the tunneling current over the plasmon widths ($\sim\tau^{-1}$) is rather independent of the choice of $\tau$ ({\it i.e.}, this property is inherited from the negligible dependence of the frequency integral of the screened interaction ${\rm Im}\{W\}$ over the plasmon width).

\begin{figure*}
\begin{center}
\includegraphics[width=1\textwidth]{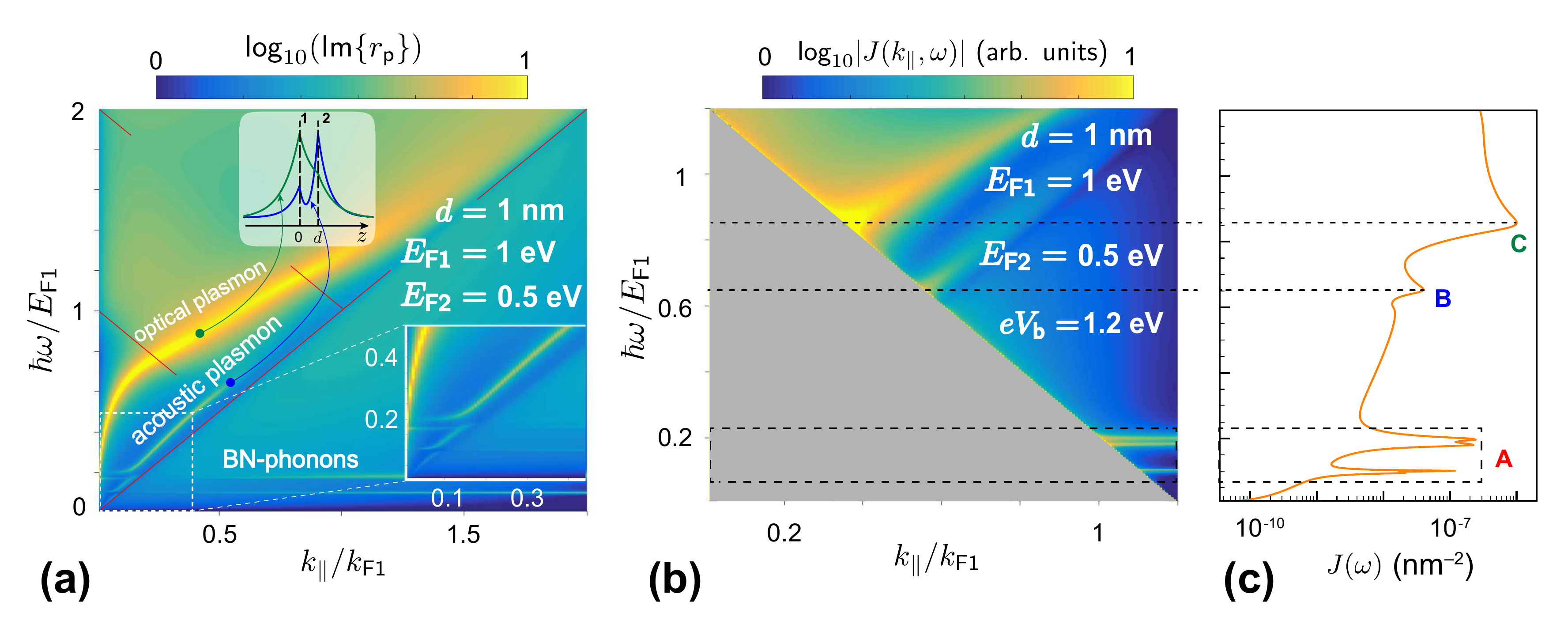}
\caption{{\bf Energy- and momentum-resolved electron tunneling}. {\bf (a)} Optical dispersion diagram of a graphene-hBN-graphene structure (Fig.\ \ref{Fig1}a), plotted through the imaginary part of the Fresnel reflection coefficient for p polarization and incidence on layer 1 as a function of optical energy $\omega$ and parallel wave-vector $\kp$. The upper inset shows the electric intensity profiles associated with acoustic and optical plasmons at the energies plotted in (c). The energy and wave-vector axes are normalized to the Fermi energy and wave vector of layer 1, $\EF_1$ and $\kf_1=\EF_1/\hbar\vf$, respectively. {\bf (b)} Contribution of different points in the dispersion diagram to inelastic electron tunneling $J(\kp,\omega)$ (Eq.\ (\ref{Gkw})). {\bf (c)} Frequency-dependent momentum-integrated tunneling probability per unit area (horizontal scale) as a function of inelastically transferred energy $\hbar\omega$ (vertical scale). We consider a graphene spacing $d=1\,$nm, Fermi energies $\Ef_1=1\,$eV and $\Ef_2=0.5\,$eV, and a bias $e\Vb=1.2\,$eV in all cases. The graphene conductivity is modeled in the RPA.}
\label{Fig2}
\end{center}
\end{figure*}

\section{Results and discussion}

\subsection{Plasmons in the Double-Layer Graphene Structure}

The optical response of the sandwich structure in Fig.\ \ref{Fig1}a is dominated by graphene plasmons and optical phonons of hBN. The frequency- and wave-vector-dependence of these excitations is illustrated in the dispersion diagram of Fig.\ \ref{Fig2}a, which shows the imaginary part of the Fresnel coefficient for p polarization and incidence from layer 1 (see Eq.\ (\ref{rp})) with a representative choice of interlayer distance $d=1\,$nm and graphene Fermi energies $\EF_1=1\,$eV and $\EF_2=0.5\,$eV. Given the small thickness of the structure, the s polarization coefficient (not shown) takes negligible values, further confirming the quasistatic behavior of the system. We consider asymmetric doping of the graphene layers, which consequently display two different plasmon bands. Similar to the hybridization observed for the surface plasmons on either side of a thin metal film \cite{PSV1975}, these bands interact and repel each other, thus pushing the upper (optical) plasmon toward higher energies and the lower (acoustic) one further down. Interestingly, nonlocality has a strong influence on the the optical response of this system: the shift produced by hybridization is limited by repulsion from the region of intraband electron-hole-pair transitions (area below the diagonal of the main plot in Fig.\ \ref{Fig2}a). As a result of this, the acoustic plasmon stays slightly above this region and exhibits nearly linear dispersion.

The gap regions free from both intra- and interband transitions (triangles defined by the downward straight lines and the main diagonal in Fig.\ \ref{Fig2}a) are different for the two graphene layers because $\EF_1\neq\EF_2$. In the smaller gap region (1\,eV maximum energy), the optical and acoustic plasmons are both well defined, with their lifetimes roughly equal to the assumed intrinsic value of $\tau$ (see above). The acoustic plasmon then fades away when its energy increases and it enters the region of interband transitions of the low-doping layer (Landau damping). A similar behavior is observed for the optical plasmons, although the described interband absorption effect is significantly weaker. Examination of the field intensity profiles associated with both plasmons explains this different behavior (Fig.\ \ref{Fig2}a, upper inset): the acoustic plasmon has larger weight on the low-doping layer (2, at $z=d$), while the optical plasmon receives substantial contributions from both layers; therefore, the effect of interband Landau damping produced by layer 1 (at $z=0$) acts more strongly on the acoustic plasmon. Additionally, optical plasmons have larger Drude weights than acoustic plasmons ({\it i.e.}, they are comparatively more weighted on layer 1, which has higher doping electron density), and this contributes to reduce the relative effect produced by coupling to the interband transitions of layer 2.

\begin{figure*}
\begin{center}
\includegraphics[width=1\textwidth]{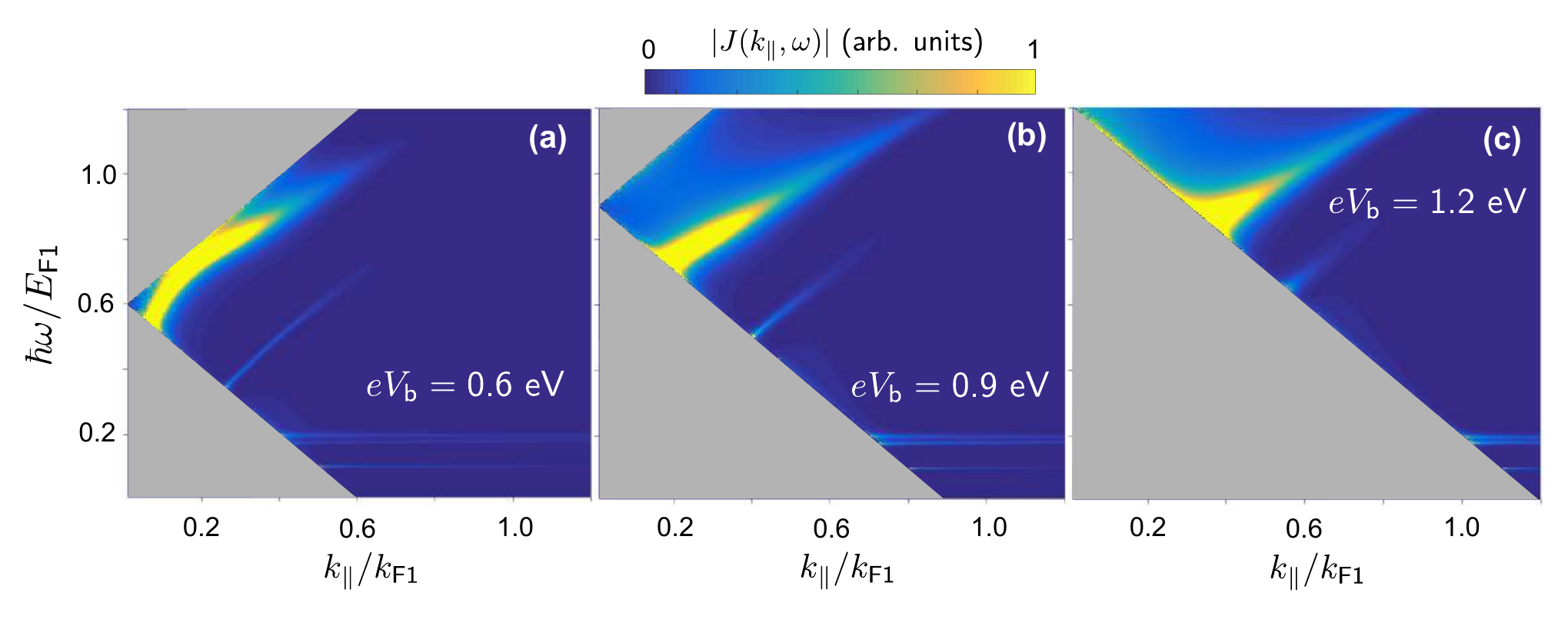}
\caption{{\bf Dispersion diagram of tunneling current components}. Contribution of different points in the dispersion diagram to the inelastic electron tunneling $J(\kp,\omega)$ (Eq.\ (\ref{Gkw})), plotted in linear scale for different bias voltages $\Vb$ (see labels) under the same conditions as in Fig.\ \ref{Fig2} ($d=1\,$nm, $\EF_1=1\,$eV, and $\EF_2=0.5\,$eV). Panel (c) is a replot of Fig.\ \ref{Fig2}b in linear scale.} 
\label{Fig3}
\end{center}
\end{figure*}

\subsection{Inelastic Tunneling Probability}

The tunneling current density can be decomposed into the contributions of different parallel-wave-vector and frequency transfers $J(\kp,\omega)$ as \[J=\int_0^\infty\! d\omega\! \int_0^\infty d\kp\, J(\kp,\omega).\]
We find that $J(\kp,\omega)$ (Eq.\ (\ref{Gkw})) exhibits features that follow the hybrid modes of the system ({\it cf.} Figure\ \ref{Fig2}a and \ref{Fig2}b). The relative strengths of the different excitations obviously depend on the applied bias voltage. Additionally, the plot of $J(k_\parallel,\omega)$ is dominated by the condition imposed by energy and momentum conservation (see Eq.\ (\ref{Gkw}) in the Appendix), which renders a zero contribution outside the region $e V_{\rm b}/\hbar+v_{\rm F}k_\parallel<\omega<e V_{\rm b}/\hbar-v_{\rm F}k_\parallel$. This is more clearly illustrated in Figure\ 3, where we show $|J(k_\parallel,\omega)|$ for different bias voltages, plotted now in linear scale. The momentum-integrated spectral decomposition of the tunneling current $J(\omega)=\int_0^\infty d\kp\, J(\kp,\omega)$ (Fig.\ \ref{Fig2}c) is thus peaked at frequencies $\omega\approx e\Vb/\hbar-\vf\kp$ for values of $\kp$ determined by the dispersion relations of the hBN phonon and the graphene plasmons.

\begin{figure*}
\begin{center}
\includegraphics[width=0.8\textwidth]{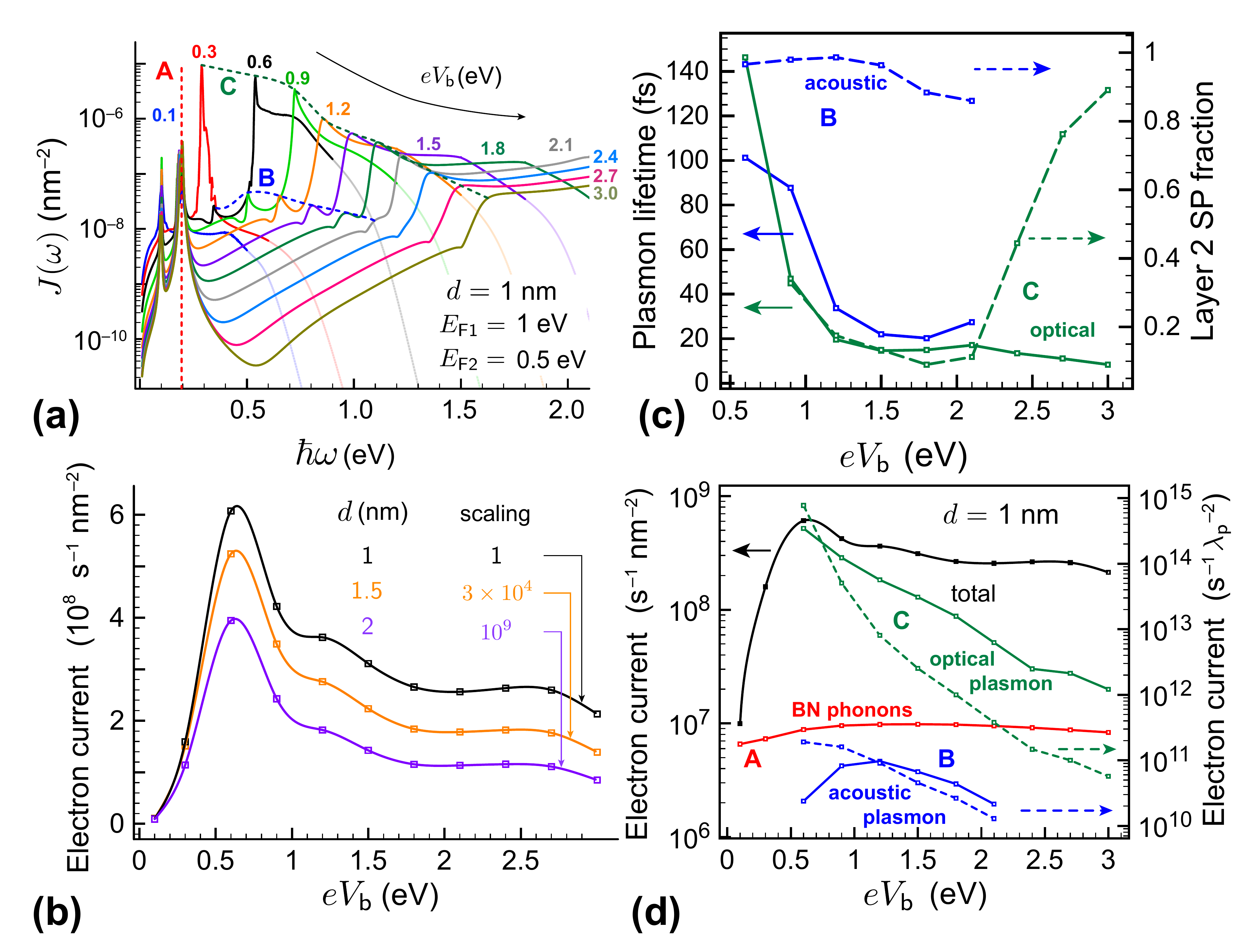}
\caption{{\bf Probability of plasmon-generation by electron tunneling}. {\bf (a)} Spectrally resolved inelastic tunneling probability for different bias voltages. {\bf (b)} Bias voltage dependence of the total inelastic tunneling current density for various graphene spacings (see labels), multiplied by the indicated factors for the sake of clarity. {\bf (c)} Plasmon lifetimes (solid curves, left scale, assuming an intrinsic RPA input lifetime of 66\,fs) and fractions of the plasmon weight in layer 2 (dashed curves, right scale, showing the fraction of squared plasmon-induced charge density in that layer). {\bf (d)} Contribution of the BN main optical phonon (A), the acoustic graphene plasmon (B), and the optical graphene plasmon (C) to the total tunneling current density (black curve), corresponding to the spectral features labeled in (a), plotted per square nanometer (solid curves, left axis) and per square plasmon wavelength (dashed curves, right axis). We consider Fermi energies of 1\,eV and 0.5\,eV for the graphene layers 1 and 2 in all cases, while the graphene spacing and bias voltage are indicated by axis labels and text insets.} 
\label{Fig4}
\end{center}
\end{figure*}

The above analysis indicates that the peak frequency of the excited plasmons can be controlled through the applied voltage. This is corroborated by Fig.\ \ref{Fig4}a, in which the peak intensities are in excellent agreement with the combination of threshold and mode dispersion relations, as we argue above. Importantly, the bias also affects the magnitude of the tunneling current (Fig.\ \ref{Fig4}a).

We are interested in producing a substantial tunneling current, but ultimately, we need that a sizable fraction of the tunneled electrons really generate plasmons. With a suitable choice of the bias voltage, the tunneling current is dominated by the production of either acoustic or optical plasmons (see Fig.\ \ref{Fig4}d). As expected, for small voltages, below the thresholds for generation of both acoustic and optical plasmons, phonons dominate the inelastic current, while for large voltages the plasmon contribution becomes small compared with other channels of inelastic tunneling associated with interband electron transitions. Therefore, there is an optimum voltage range in which the generation yield (plasmons per electron) has unity order; tunneling in that range is actually dominated by the excitation of optical plasmons (see region near $eV_{\rm b}\sim\EF_1=1\,$eV in Fig.\ \ref{Fig4}d).

Additionally, we are interested in producing long-lived plasmons. Under the conditions of Fig.\ \ref{Fig4}, for a realistic bias voltage $<1\,$V$/$nm, the lifetimes are boosted when the plasmons enter the gap region of the low-doping graphene layer 2 (Fig.\ \ref{Fig4}c, solid curves), where they reach values of the order of the assumed intrinsic lifetime $\tau$, which depends on material quality and is  ultimately limited by acoustic phonons \cite{WLG15}.

Similar qualitative results are obtained when considering other values of the Fermi energies or when the graphene layers are separated by a vacuum gap, as shown in the additional plots presented in the Appendix.

\section{Concluding remarks}

In order to place the calculated tunneling current densities in context, it is pertinent to consider the current produced over an area of a squared plasmon wavelength. As shown in Fig.\ \ref{Fig4}c (right scale), this quantity reaches values of $\sim10^{12}-10^{14}/$s for a realistic graphene spacing of 1\,nm, corresponding to roughly three hBN atomic layers. In this respect, the maximum applied voltage before breakdown takes place is an important factor to consider, as well as the threshold voltage for plasmon generation, which depends on the Fermi energies of the two graphene layers. It is important to stress that the generation efficiency reaches one plasmon per tunneled electron, which should enable the generation of single and few plasmons when combined with quantum electron transport setups \cite{IGM10}.

Electrical detection of graphene plasmons can also be accomplished within the sandwich structures under consideration. A plasmon propagating through the structure or laterally confined in a finite-size island can decay by transferring its energy to a tunneling electron under the appropriate bias conditions. Indeed, using high-quality graphene and considering plasmons of frequency and momentum in the gap region of both graphene layers, inelastic electron tunneling should be the dominant channel of plasmon decay, leaving an electron in the low-potential layer and a hole in the high-potential one. In this way, charge separation is naturally accomplished, thus facilitating the electrical detection of the plasmon. There are two favorable factors that support this possibility in graphene: the noted momentum mismatch that prevents direct elastic tunneling (obviously, graphene quality is an important factor to prevent defect-assisted elastic tunneling); and the high spatial concentration of the plasmons, which result in large coupling to inelastic transitions compared with coupling to photons in tunneling-based inelastic light emission measurements \cite{BGJ91}. As an alternative approach, electrical detection could be performed by a separate graphene-based structure {\it via} thermoelectric measurements \cite{LGW16} or through thermo-optically activated nanoscale junctions \cite{paper275}.

These concepts are equally applicable to other van der Waals crystals supporting 2D polaritons ({\it e.g.}, plasmons in black phosphorous \cite{LRW14,HMP17}, or even optical phonons in hBN). Then, our formalism can be easily adapted to materials other than graphene by correcting the electron energies and matrix elements of Eq.\ (\ref{Gkw}) ({\it e.g.}, using tight-binding electron wave functions \cite{LSY13}), as well as the conductivities of Eq\ (\ref{W}).

The sandwich structure under consideration is convenient for the design of integrated plasmonic circuitry. We envision plasmon generation and detection in patches with a lateral size of the order of the plasmon wavelength ($\sim 10-100\,$nm for the plasmon energies under consideration). The proposed source generates plasmons peaked at bias-dependent frequencies, but further frequency selection could be performed upon transmission of the plasmons through engineered waveguides. These elements should grant us access into optics-free devices, in which plasmons can perform different operations, for example by interacting with quantum dots and localized molecular excitations, truly relying on their very nature as collective electron excitations, without the mediation of photons whatsoever. Plasmon-based sensors could then consist of a plasmon generation stage, a transmission ribbon exposed to the analyte, and a plasmon detection stage, with some degree of spectral resolution possibly achieved by playing with the graphene Fermi levels and the applied bias voltages. The versatility and excellent performance predicted for the double-graphene sandwich structure opens exciting prospects for the design of modular integrated devices with multiple functionalities.

\appendix

\section{Calculation of Inelastic Electron Tunneling Probabilities}

We consider electron transitions between initial $i$ and final $f$ states assisted by the creation of inelastic excitations in the system. The initial and final states are taken to be in the graphene layers 1 and 2, repectively. The transition probability $\Gamma$ can be decomposed into the contribution of different inelastic frequencies $\omega$ as $\Gamma=\int_0^\infty\! d\omega\,\Gamma(\omega)$, where the spectrally resolve probability admits the expression \cite{paper149}
\begin{widetext}
\begin{align}
	\Gamma(\omega) = \dfrac{2e^2}{\hbar} \sum_{i,f} \int\! d^3\rb \int\! d^3\rb'\; &\psi_i^*(\rb) \psi_f(\rb) \psi_f^*(\rb') \psi_i(\rb') \text{Im}\{ - W(\rb,\rb',\omega) \} \label{Gamma} \\ 
	& \times \delta(\varepsilon_f - \varepsilon_i + \omega) \; f_1(\hbar \varepsilon_i) \; [1 - f_2(\hbar \varepsilon_f)].
\nonumber
\end{align}
\end{widetext}
Here, $\psi_i$ and $\psi_f$ are initial and final electron wave functions of energies $\hbar\varepsilon_i$ and $\hbar\varepsilon_f$, respectively, $f_j(E)=\{1+\exp[(E-\Ef_j)/k_\text{B}T]\}^{-1}$ is the Fermi-Dirac electron distribution of the graphene layer $j$ (we set $ T=300 $\,K), and $W(\rb,\rb',\omega)$ is the screened interaction, defined as the electric potential produced at $\rb$ by a unit charge oscillating with frequency $\omega$ at the position $\rb'$. This result is rigorous up to first-order perturbation in the screened interaction $W$. We assume that the electron wave functions of the system depicted in Fig.\ \ref{Fig1}a can be factorized as the product of decoupled in-plane and out-of-plane components. The latter, which we approximate by a real quantum-well wave function  $\phiwell(z)$ (see below), is shared by all conduction electrons. The remaining in-plane wave functions are Dirac fermions \cite{CGP09} characterized by their 2D wave vectors $\Qb$. Additionally, the translational symmetry of the system allows us to write the screened interaction as $W(\rb,\rb',\omega) = (2\pi)^{-2}\!\int\! d^2\kpb \exp[\ii\kpb\!\cdot\!(\Rb - \Rb')]\, W(\kp,z,z',\omega)$ in terms of parallel wave vector components $\kpb$. Explicit expresions for $W(\kp,z,z',\omega)$ are offered below. Likewise, the transition rate can be separated as $\Gamma=\int_0^\infty\! d\omega\! \int_0^\infty d\kp\, \Gamma(\kp,\omega)$. We then divide by the sandwich area $A$ to obtain the tunneling current $J=\Gamma/A$. As we show in the detailed derivation provided in Sec.\ \ref{Sec2}, Eq.\ (\ref{Gamma}) allows us to write the wave-vector- and frequency-resolved tunneling current as
\begin{widetext}
\begin{align}
J(\kp,\omega)=\dfrac{e^2\kp}{2\pi^3 \hbar} & \int d^2\Qb_i
\; \left( 1 + \dfrac{\Qb_i \cdot (\Qb_i-\kpb)}{Q_i \; |\Qb_i-\kpb|} \right)
\; f_1(\hbar\vf Q_i) \; [1-f_2(\hbar \vf |\Qb_i-\kpb|)] \label{Gkw}\\
& \times\; \int\! d z \! \!\int\! d z' \;
\phiwell(z)\phiwell(z-d)\phiwell(z')\phiwell(z'-d) \;
{\rm Im}\left\{-W(\kp,z,z',\omega) \right\} \nonumber\\
& \times\; \delta(\vf(|\Qb_i - \kpb| - Q_i) + \omega-e\Vb/\hbar), \nonumber 
\end{align}
\end{widetext}
where $\vf\approx10^6\,$m$/$s is the graphene Fermi velocity. This expression, which is independent of the direction of $\kpb$, includes an integral over initial-state parallel wave vectors $\Qb_i$, as well as a factor of 4 accounting for spin and valley degeneracies. Additionally, the $\delta$ function imposes energy conservation and limits the spectral range for which $J(\kp,\omega)$ is nonzero to
$eV_{\rm b}/\hbar-v_{\rm F}k_\parallel<\omega<eV_{\rm b}/\hbar+v_{\rm F}k_\parallel$


\section{Out-of-Plane Graphene Electron Wave Function}

We approximate the dependence of the graphene electron wave function $\phiwell(z)$ on the out-of-plane direction $z$ by one of the states of a quantum well. More precisely, we consider the first excited well state, in order to mimic the antisymmetry of the $p_z$ orbitals in graphene conduction band. For simplicity, we assume a square well potential of depth $V_0$ and width $a$. These parameters are fitted in such a way that {\it (i)} the binding energy of the well state coincides with the graphene work function $\Phi=4.7 $\,eV \cite{YZR09}; and {\it (ii)} the centroid of the electron-density spill-out is the same in the well state and in the $p_z$ orbital (i.e., $\int\! dz\,|z|\,|\phiwell(z)|^2=\int\! d^3\rb\,|z|\,|\varphi_{p_z}(\rb)|^2$). Using the atomic carbon $2p$ orbital for $\varphi_{p_z}$ \cite{CR1974}, we find $V_0=45 $\,eV and $a=0.12\,$nm. This value of $a$ is close to (but smaller than) the interlayer spacing in graphite (0.335\,nm), which is reasonable due to the spill out of $\phiwell$ outside the well. A detailed derivation of these results is given in Sec.\ \ref{Sec1}. Incidentally, the graphene-electron spill-out toward the gap must depend on the details of the electron band structure, which should also vary with the filling material. In particular, hBN presents a band gap of $\sim6\,$eV \cite{WTK04}, so the jump between its conduction-band bottom and the chemical potential is similar to the work function of graphene, and therefore, for simplicity we ignore band-structure effects in hBN and calculate the tunneling electron states by assuming a gap potential at the level of the surrounding vacuum.

\section{Screened Interaction}

We consider a sandwich formed by two graphene layers of conductivities $\sigma_1$ and $\sigma_2$, separated by an anisotropic film of dielectric permittivity $\epsilon_z$ and $\epsilon_x$ for directions parallel ($z$) and perpendicular ($x$, $y$) to the film normal, respectively, and surrounded by vacuum above and below the structure. In the electrostatic limit here assumed, the surface-normal wave vector inside the dielectric can be written as $\ii q$ with $q=\kp\sqrt{\epsilon_x/\epsilon_z}$, where we take the sign of the square root such that ${\rm Re}\{q\}>0$. It is also convenient to define the effective permittivity $\epsilon=\sqrt{\epsilon_x\epsilon_z}$, taken to have positive imaginary part. The screened interaction admits the closed-form expression (see Appendix)
\begin{widetext}
\begin{align}
W(\kp,z,z',\omega)=W^{\rm dir}(\kp,z,z',\omega)+W^{\rm ref}(\kp,z,z',\omega),
\label{W}
\end{align}
where
\begin{align}
W^{\rm dir}(\kp,z,z',\omega)=\frac{2\pi}{\kp}\times
\left\{\begin{array}{ll}
\ee^{-\kp|z-z'|},  &z,z'<0 \text{ or } z,z'>d \\
\epsilon^{-1}\ee^{-q|z-z'|},  &0<z,z'<d \\
0, & \text{otherwise}
\end{array} \right.
\nonumber
\end{align}
is the interaction between charges placed inside the homogeneous vacuum or dielectric parts of the structure, while
\begin{align}
&W^{\rm ref}(\kp,z,z',\omega)=\frac{(2\pi/\kp)}{1-A'_1A'_2\ee^{-2qd}}
\nonumber\\
&\times
\left\{\begin{array}{lll}
\ee^{\kp(2d-z-z')}\left[A_2+A'_1(A_2+B'_2)\,\ee^{-2qd}\right],  &d<z &d<z' \\ 
B_2\;\ee^{\kp(d-z)}\left[\ee^{-q(d-z')}+A'_1\;\ee^{-q(d+z')}\right],  & d<z, &0<z'<d \\ 
\epsilon B_1B_2\;\ee^{\kp (d-z+z')}\ee^{-qd},  & d<z, &z'<0 \\ 
B_2\;\ee^{\kp(d-z')}\left[\ee^{-q(d-z)}+A'_1\;\ee^{-q(d+z)}\right],  & 0<z<d, &d<z' \\ 
\epsilon^{-1}\big\{
A'_1\;\ee^{-q(z+z')}+A'_2\;\ee^{-q(2d-z-z')} & & \nonumber\\
\quad\;\;\;
+A'_1A'_2\left[\ee^{-q(2d+z-z')}+\ee^{-q(2d-z+z')}\right]
\big\},  & 0<z<d, &0<z'<d \\ 
B_1\;\ee^{\kp z'}\left[\ee^{-qz}+A'_2\;\ee^{-q(2d-z)}\right],  & 0<z<d, &z'<0 \\ 
\epsilon B_1B_2\;\ee^{\kp(d+z-z')}\ee^{-qd},  &z<0, &d<z'\\ 
B_1\;\ee^{\kp z}\left[\ee^{-qz'}
+A'_2\;\ee^{-q(2d-z')}\right],  &z<0, &0<z'<d\\ 
\ee^{\kp(z+z')}\left[A_1+A'_2(A_1+B'_1)\,\ee^{-2qd}\right],  &z<0, &z'<0
\end{array} \right.
\nonumber
\end{align}
\end{widetext}
is the contribution produced by reflections at the graphene layers. We have used in this expression transmission and reflection coefficients of the individual dielectric/graphene/vacuum interface for the electric potential defined by $B_j=2/(1+\epsilon+\beta_j)$, $B'_j=\epsilon B_j$, $A_j=B_j-1$, and $A'_j=B'_j-1$, where $\beta_j=4\pi\ii\kp\sigma_j/\omega$. Reassuringly, Eq.\ (\ref{W}) explicitly satisfies the reciprocity condition $W(\kp,z,z',\omega)=W(\kp,z',z,\omega)$. In our calculations, we consider two graphene layers separated by either vacuum ($\epsilon_x=\epsilon_z=1$) or hBN. For hBN, we approximate optical phonons as Lorentzians and write \cite{GPR1960}
$\epsilon_{\ell} (\omega)\approx\epsilon_{\infty,\ell}+\sum_{i=1,2}s_{\ell i}^2/[\omega_{\ell i}^2-\omega(\omega+\ii\gamma_{\ell i})]$, 
with parameters
$\epsilon_{\infty,z}=4.10$,
$s_{z 1}=70.8\,$meV,
$\omega_{z 1}=97.1\,$meV,
$\gamma_{z 1}=0.99\,$meV,
$s_{z 2}=126\,$meV,
$\omega_{z 2}=187\,$meV,
and
$\gamma_{z 2}=9.92\,$meV
for the out-of-plane component ($\ell=z$); and
$\epsilon_{\infty,x}=4.95$,
$s_{x 1}=232\,$meV,
$\omega_{x 1}=170\,$meV,
and
$\gamma_{x 1}=3.6\,$meV,
$s_{x 2}=43.5\,$meV,
$\omega_{x 2}=95.1\,$meV,
and
$\gamma_{x 2}=3.4\,$meV
for the in-plane component ($\ell=x$). The graphene conductivities $\sigma_j(\kp,\omega)$ are calculated in the RPA \cite{W1947,WSS06,HD07}, including nonlocal effects through their $\kp$ dependence. We assume a graphene plasmon lifetime $\tau=66\,$fs ({\it i.e.}, $\hbar\tau^{-1}=10\,$meV) in all cases.

\section{Fresnel's Reflection Coefficient}

Within the quasistatic approximation, the Fresnel reflection coefficient of the graphene-hBN-graphene sandwich vanishes for s polarization, whereas it can be readily extracted from Eq.\ (\ref{W}) for p polarization. Indeed, considering a distant point source in the $z<0$ region, the external and reflected scalar potentials below the sandwich are $\propto (2\pi/\kp)\ee^{-\kp z}$ and $\propto -r_{\rm p}\;(2\pi/\kp)\ee^{\kp z}$, which allow us to identify
\begin{equation}
r_{\rm p}=-\frac{A_1+A'_2(A_1+B'_1)\ee^{-2qd}}{1 - A'_1A'_2\ee^{-2qd}}
\label{rp}
\end{equation}
by comparison to Eq.\ (\ref{W}) for light incident on layer 1.

\section{Effective quantum-well description of the out-of-plane graphene electron wave function}
\label{Sec1}

We describe the out-of-plane electron wave function of conduction electrons in graphene as a quantum well state. For simplicity, we assume a square well potential of depth $V_0$ and width $a$. We focus on the first excited state, which is antisymmetric along the normal direction $z$, just like the $p_z$ orbital in graphene, as shown in the following scheme:
\\ \begin{center}
\includegraphics[width=0.35\textwidth]{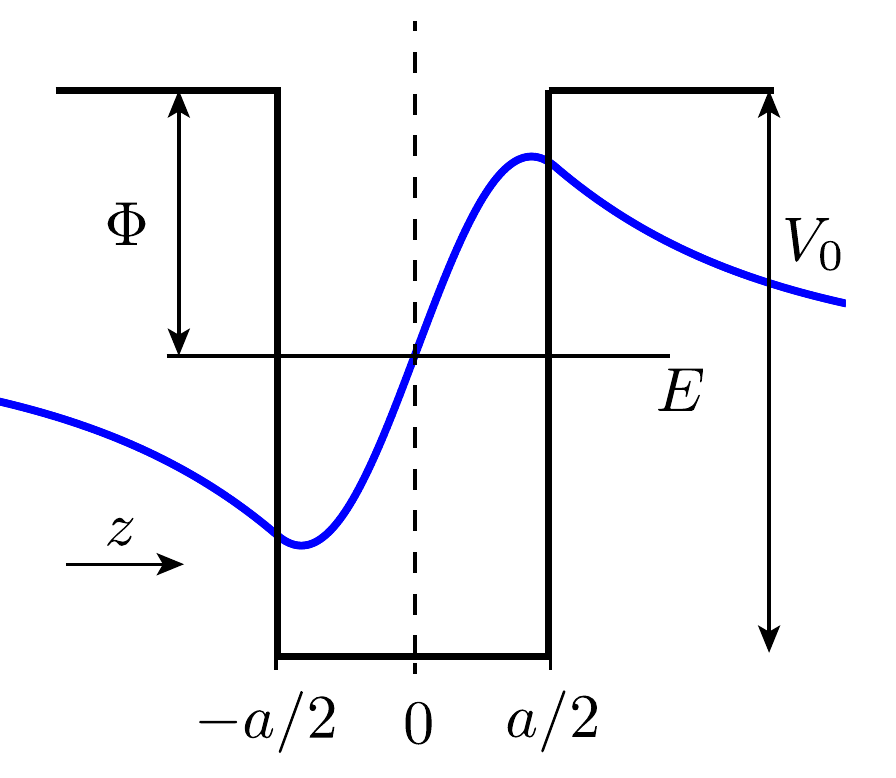}
\end{center}
We are therefore interested in the red wave function, which we denote $\phiwell(z)$. In this model, the parameters $V_0$ and $a$ are used to fit {\it (i)} the state binding energy to the graphene work function \cite{YZR09} $\Phi=4.7$\,eV; and {\it (ii)} the centroid of the electron density outside the $z=0$ plane to the value obtained for the carbon $2p$ orbital.

We find the quantum well state $\phiwell(z)$ by solving the Schr\"odinger equation $(-\hbar^2/2m)d^2\phiwell(z)/dz^2+[V(z)-E] \phiwell(z)=0$, where $V(z)$ is the potential shown in the scheme above. The wave function of the first excited state can be written $A\sin(k_{\rm in}z)$ inside the well, with $k_{\rm in}=\sqrt{2m(E+V_0)}/\hbar$, while it decays evanescently in the outer region as ${\rm sign}(z) B\ee^{-k_{\rm out}|z|}$, with $k_{\rm out}=\sqrt{-2mE}/\hbar$. The electron energy $E<0$ is referred to the potential outside the well. The continuity of the wave function and its derivate at the well boundaries $z=\pm a/2$ lead to the condition $-k_{\rm in}/k_{\rm out}=\tan(k_{\rm in}a/2)$, which determines the discrete energies $E$ of asymmetric states. Combining these conditions with normalization ($\int dz |\phiwell(z)|^2=1$), we find $|A|^2=(a/2+1/k_{\rm out})^{-1}$ and $|B|^2=|A|^2\,\ee^{k_{\rm out}a} \sin^2(k_{\rm in}a/2)$.

The centroid of the wave function away from the $z=0$ plane is calculated as $\int\!dz\, |z|\,|\phiwell(z)|^2$ and compared with the centroid of the $p_z$ orbital of graphene $\int\!d^3\rb\,|z|\,|\varphi_{p_z}(\rb)|^2$. We approximate the latter by using a tabulated $2p$ atomic carbon wave function \cite{CR1974}, $\varphi_{2p}(\rb)=z\sum_j \beta_j e^{-\alpha_j r}$, where the parameters $\alpha_j$ and $\beta_j$ are expressed in the following table in atomic units:
\begin{center}
\begin{tabular}{ c  c  c}
$j$ & $\alpha_j$ & $\beta_j$ \\ \hline
1 & \quad 1.10539 \quad & \quad 0.4610 \quad \\
2 & \quad 0.61830 \quad & \quad 0.0134 \quad \\
3 & \quad 2.26857 \quad & \quad 1.5905 \quad \\
4 & \quad 5.23303 \quad & \quad 0.7291 \quad \\
\end{tabular}
\end{center}
In the following plot we show the $2p$ electron probability density integrated over parallel $(x,y)$ directions ($\int\! dx \int\! dy \; |\varphi_{2p} (\rb)|^2$, solid curve), compared with the fitted well state ($|\phiwell(z)|^2$, dashed curve):
\\ \begin{center}
\includegraphics[width=0.45\textwidth]{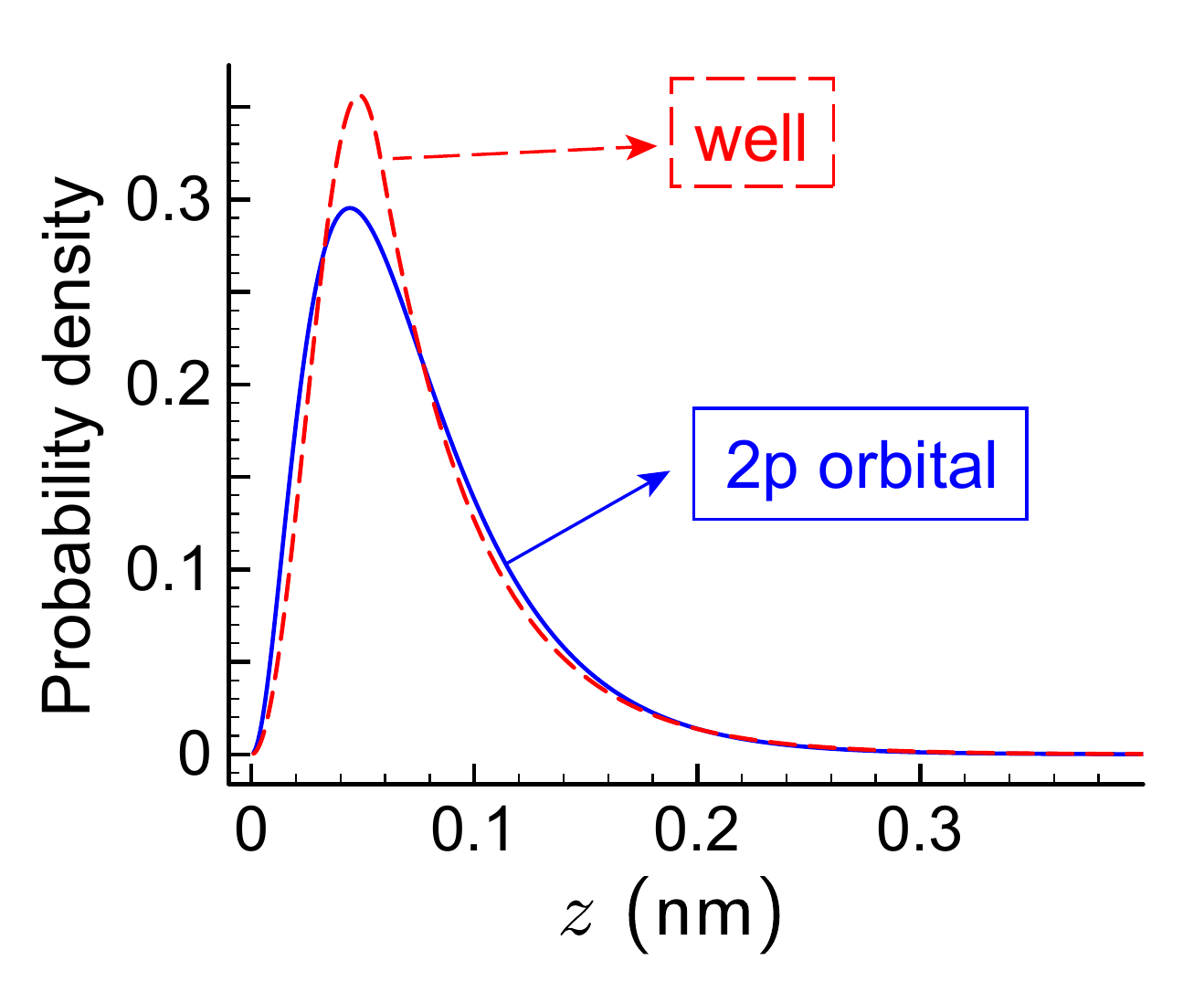}
\end{center}
The agreement between the two probability densities is excellent using fitted values $V_0=45\,$eV and $a=0.12\,$nm.

\section{Derivation of Eq.\ (\ref{Gkw})}
\label{Sec2}

In this section, we start from Eq.\ (\ref{Gamma}) for the inelastic electron transition probability,
\begin{widetext}
\begin{align}
\Gamma(\omega) = \dfrac{2e^2}{\hbar} \sum_{i,f} \int\! d^3\rb \int\! d^3\rb' \; &\psi_i^\dagger(\rb)\cdot\psi_f(\rb) \; \psi_f^\dagger(\rb')\cdot\psi_i(\rb') \; \text{Im}\{ - W(\rb,\rb',\omega) \} \label{Gamma} \\ 
& \times \delta(\varepsilon_f - \varepsilon_i + \omega) \; f_1(\hbar \varepsilon_i) \; [1 - f_2(\hbar \varepsilon_f)]
\nonumber
\end{align}
\end{widetext}
(see definitions of different elements in the Appendix), and specify it for the sandwich structure depicted in the following scheme, consisting of two graphene layers separated by a film of thickness $d$ and permittivity $\epsilon$:
\\ \begin{center}
\includegraphics[width=0.35\textwidth]{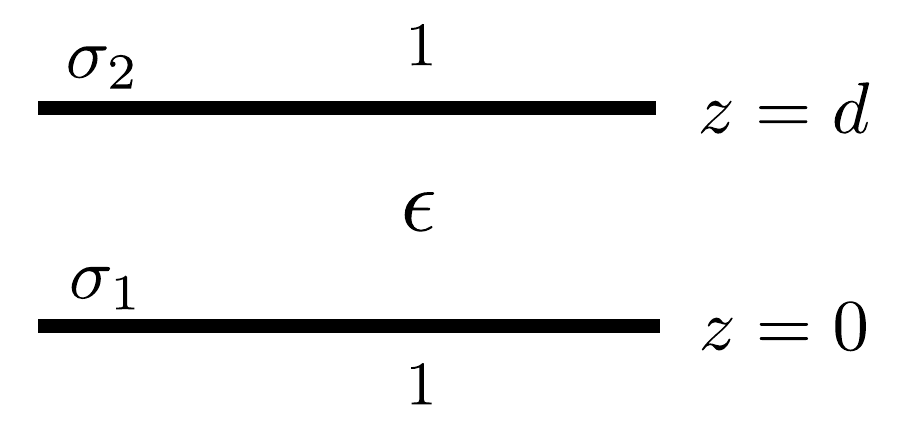}
\end{center}
We factorize the electron wave functions as the product of in-plane $\varphi^\parallel$ and out-of-plane $\phiwell$ components. The latter is described in Sec.\ \ref{Sec1} and is common to all conduction electrons, so we can write $\psi_i(\rb)=\varphi_i^\parallel(\Rb)\phiwell(z)$ for initial states in layer 1, centered at $z=0$, and $\psi_f(\rb)=\varphi_f^\parallel(\Rb)\phiwell(z-d)$ for final states in layer 2, centered at $z=d$. Here, we use the notation $\rb=(\Rb,z)$, with $\Rb=(x,y)$.

The in-plane wave functions are Dirac fermions characterized by their parallel wave vector $\Qb=(Q_x,Q_y)$, spin, and valley (K or K' points). The transition probability must be independent of spin and valley, so we perform the calculation for only one combination of these degrees of freedom and multiply the result by a factor of 4. We further consider negative doping and a bias such that the initial and final wave functions lie within the upper (conduction) band of their respective layers. The Dirac fermions admit the expression \cite{CGP09}
\begin{equation}
\varphi^\parallel(\Rb) = \dfrac{1}{\sqrt{2A}} \ee^{\ii \Qb\cdot\Rb} \binom{\ee^{\ii \phi_\Qb /2}}{\ee^{-\ii \phi_\Qb /2}},
\label{eq:wfg}
\end{equation}
where $A$ is the normalization area, $\phi_\Qb= \tan^{-1}(Q_y/Q_x)$ is the azimuthal angle of $\Qb$, and the upper and lower components refer to the value of the wave function in each of the two graphene carbon sublattices. The corresponding electron energy relative to the Dirac point is $\hbar\varepsilon_\Qb=\hbar \vf Q$.

We have adapted Eq.\ (\ref{Gamma}) from a previous derivation \cite{paper149} in order to incorporate the sum over both carbon sublattices, which is accounted for through the indicated spinor products. Before inserting Eq.\ (\ref{eq:wfg}) into Eq.\ (\ref{Gamma}), we recast the sum over $i$ as $\sum_i\rightarrow (A/4\pi^2)\int d\Qb_i$, and similarly for the sum over $f$. Additionally, as a consequence of translational invariance, the integrand inside $\int\! d^3\rb$ should be independent of $\Rb$, so we can replace $\int\! d^2\Rb\rightarrow A$. Now, we express the screened interaction as $W(\rb,\rb',\omega) = (2\pi)^{-2}\!\int\! d^2\kpb \exp[\ii\kpb\!\cdot\!(\Rb - \Rb')]\, W(\kp,z,z',\omega)$ (see Sec.\ \ref{Sec3}), which allows us to carry out the integral over $\Rb'$ analytically to yield a $\delta$ function for conservation of parallel momentum, $\delta(\Qb_f-\Qb_i+\kpb)$, and this in turn can be used to perform the integral over $\Qb_f$. Putting these elements together, Eq.\ (\ref{Gamma}) becomes
\begin{widetext}
\begin{align}
\Gamma(\omega) = \dfrac{e^2A}{8\pi^4\hbar} &\int d^2\kpb \int d^2\Qb_i \;
\left|(\ee^{-\ii \phi_{\Qb_f}/2}\quad\ee^{\ii \phi_{\Qb_f}/2})\cdot\binom{\ee^{\ii \phi_{\Qb_i}/2}}{\ee^{-\ii \phi_{\Qb_i}/2}}\right|^2
\; f_1(\hbar\vf Q_i) \; [1-f_2(\hbar \vf |\Qb_i-\kpb|)] \nonumber\\
& \times\; \int\! d z \! \!\int\! d z' \;
\phiwell(z)\phiwell(z-d)\phiwell(z')\phiwell(z'-d) \;
{\rm Im}\left\{-W(\kp,z,z',\omega) \right\} \nonumber\\
& \times\; \delta(\vf(|\Qb_i - \kpb| - Q_i) + \omega-e\Vb/\hbar). \label{Gammabis} 
\end{align}
\end{widetext}
Notice that the Fermi-Dirac distributions are referred to the Dirac point of their respective graphene layers. However, the electron energy in layer 2 is shifted by the bias energy $-e\Vb$ relative to layer 1 (last term inside the $\delta$ function). We also note that the spinor product yields $\left|1+\exp[\ii(\phi_{\Qb_i}-\phi_{\Qb_f})]\right|^2=2\left[1+\Qb_i \cdot(\Qb_i-\kpb) Q_i^{-1}|\Qb_i-\kpb|^{-1}\right]$, where we have expressed the angle between $\Qb_i$ and $\Qb_f=\Qb_i-\kpb$ in terms of the inner product of these two vectors. Finally, inserting this expression into Eq.\ (\ref{Gammabis}), and noticing that the result is independent of the direction of $\kpb$ once the $\Qb_i$ integral has been carried out, we can make the substitution $\int\!d\phi_\kpb\rightarrow2\pi$ for the azimuthal integral and divide the result by the graphene area $A$ to readily obtain Eq.\ (2) of the main text. Equation\ 2 is the expression that we use in our numerical simulations of the tunneling current, in which the azimuthal $\phi_{\Qb_i}$ integral is carried out analytically by using the relation $\delta[F(\phi_{\Qb_i})]=\sum_j \delta(\phi_{\Qb_i}-q_j)/|F'(q_j)|$ for the $\delta$ function (notice that the poles of $F(\phi_{\Qb_i})$ are of first order).

\section{Derivation of Eq.\ (\ref{W})}
\label{Sec3}

The screened interaction $W(\rb,\rb',\omega)$ is defined as the scalar potential produced at $\rb$ by a charge placed at $\rb'$ and oscillating with frequency $\omega$. Translational invariance allows us to write
\begin{equation}
W(\rb,\rb',\omega)=\int\! \dfrac{d^2\kpb}{(2\pi)^2}  \ee^{\ii \kpb\cdot(\Rb - \Rb')} W (\kp,z,z',\omega),
\nonumber
\end{equation}
so it is natural to work in $\kpb$ space ({\it i.e.}, we assume an overall $\ee^{\ii\kpb\cdot(\Rb-\Rb')}$ dependence). A point charge placed at $z'$ produces a direct scalar potential $(2\pi/\kp)\ee^{-\kp|z-z'|}$ in vacuum ({\it i.e.}, this is the direct Coulomb interaction term in Eq.\ (3)). Additionally, inside the bulk of an anisotropic dielectric (permittivity $\epsilon_z$ along $z$, and $\epsilon_x$ along $x$ and $y$), the Poisson equation $\nabla\cdot\epsilon\nabla\phi=0$ has solutions $\phi=\ee^{\pm\ii qz}$, where $q=\kp\sqrt{\epsilon_x/\epsilon_z}$ and we take the square root to yield ${\rm Im}\{q\}>0$; this allows us to write the point-charge potential as $[2\pi/(\epsilon_z q)]\ee^{-q|z-z'|}$ inside that medium. Now, the induced potential has the form $A\ee^{\kp z}$ below the sandwich ($z<0$), $D\ee^{\kp(d-z)}$ above it ($z>d$), and $B\ee^{-qz}+C\ee^{-q(d-z)}$ inside the dielectric ($0<z<d$). Here, the coefficients $A$, $B$, $C$, and $D$ are used to satisfy the boundary conditions, namely: (1) the continuity of the potential at each graphene layer $j=1,2$; and (2) the jump of normal displacement is equal to $4\pi$ times the induced charge. From the continuity equation, the induced charge can be expressed as the divergence of the current, and this in turn as the product of the conductivity $\sigma_j$ times the in-plane electric field. The jump of normal displacement at layer $j$ is then given by $-4\pi\ii k_\parallel^2\sigma_j/\omega$ times the potential. Solving the resulting system of four equations for each position of the external charge $z'$, we obtain, after some tedious but  straightforward algebra, the expression for the screened interaction $W (\kp,z,z',\omega)$ presented in Eq.\ (3) of the main text. Alternatively, a more direct Fabry-Perot-like derivation can be made in terms of the transmission and reflection coefficients of the dielectric/graphene/vacuum interface defined in the main text.

\begin{figure}
\begin{center}
\includegraphics[width=0.5\textwidth]{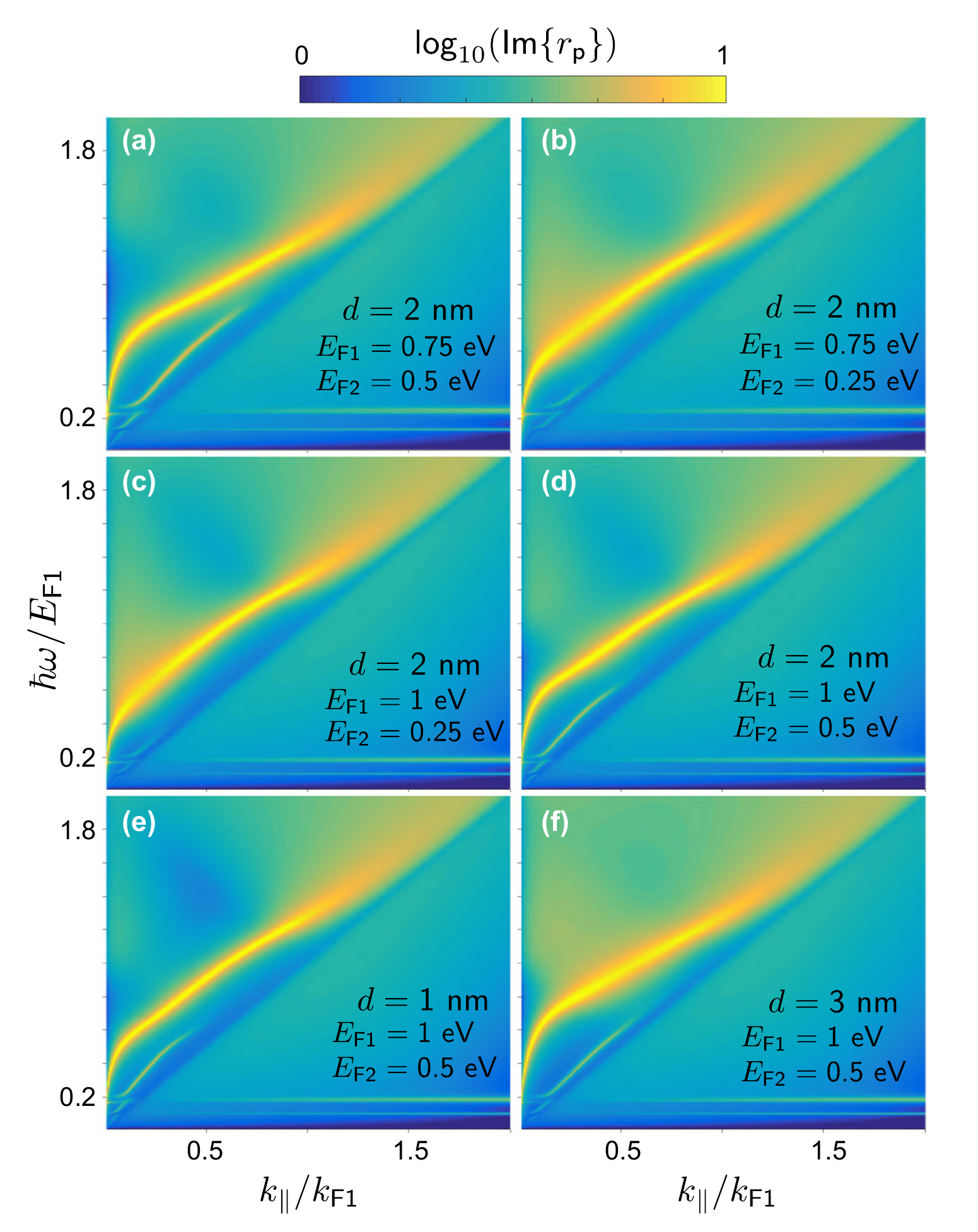}
\caption{{\bf Additional dispersion diagrams.} Same as Fig.\ \ref{Fig2}a for various combinations of the gap distance and the graphene Fermi energies (see labels).}
\label{FigS1}
\end{center}
\end{figure}

\begin{figure}
\begin{center}
\includegraphics[width=0.5\textwidth]{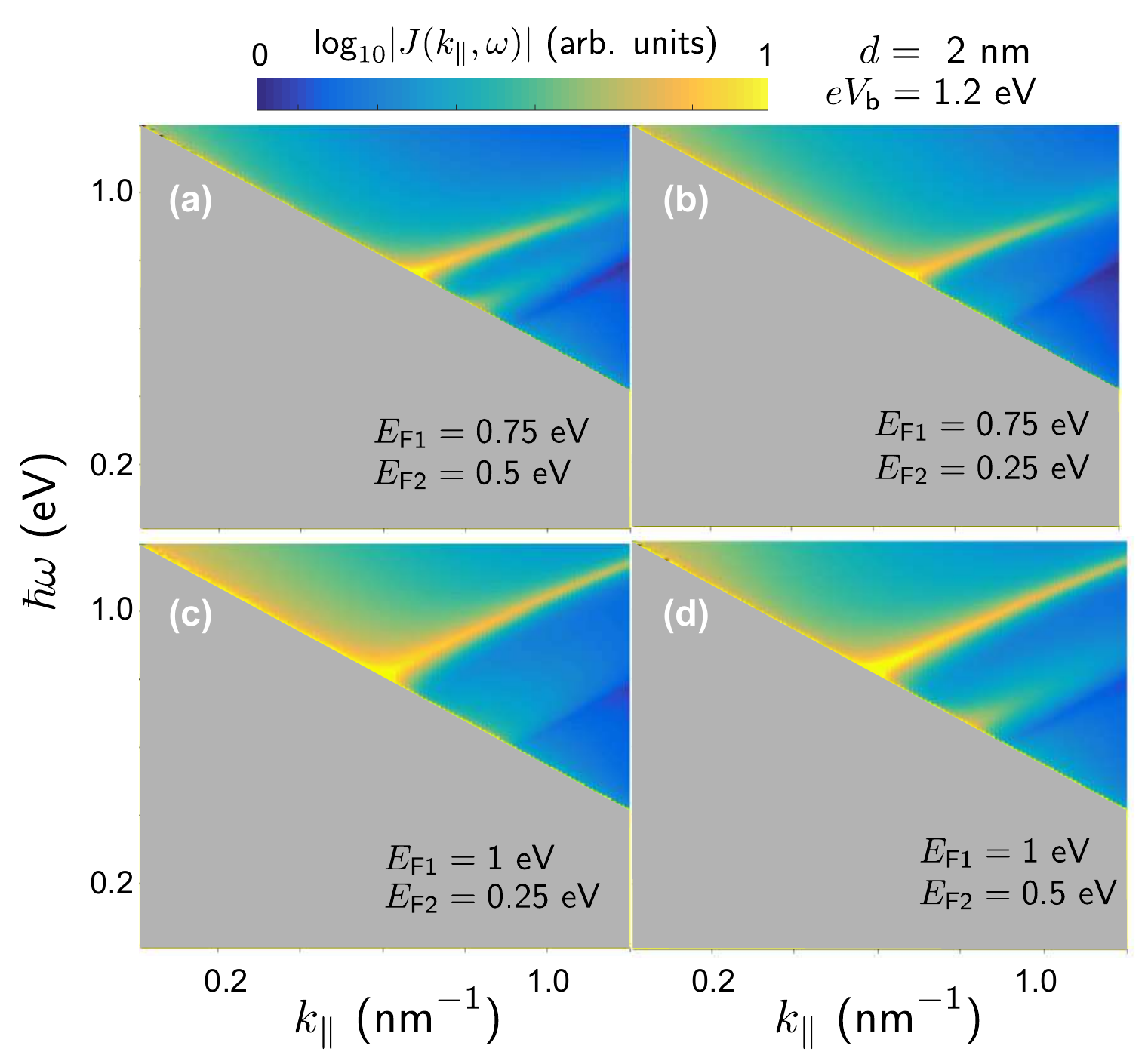}
\caption{{\bf Additional calculations of the energy- and momentum-resolved tunneling current.} Same as Fig.\ \ref{Fig2}b for fixed gap distance $d=2\,$nm and various combinations of the graphene Fermi energies (see labels).}
\label{FigS2}
\end{center}
\end{figure}

\begin{figure*}
\begin{center}
\includegraphics[width=1\textwidth]{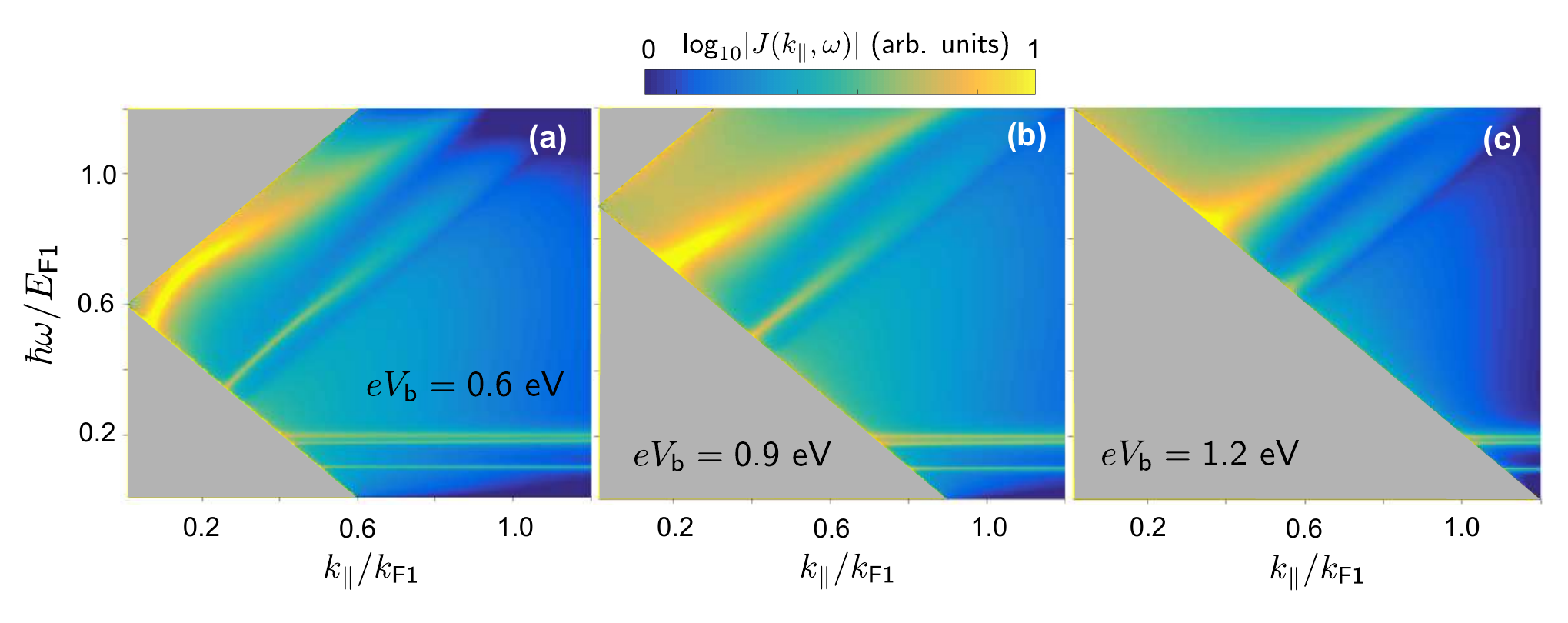}
\caption{{\bf Additional dispersion current diagrams.} Same as Fig.\ \ref{Fig3} with $|J(\kp,\omega)|$ plotted in logarithmic scale.}
\label{FigS3}
\end{center}
\end{figure*}

\begin{figure}
\begin{center}
\includegraphics[width=0.5\textwidth]{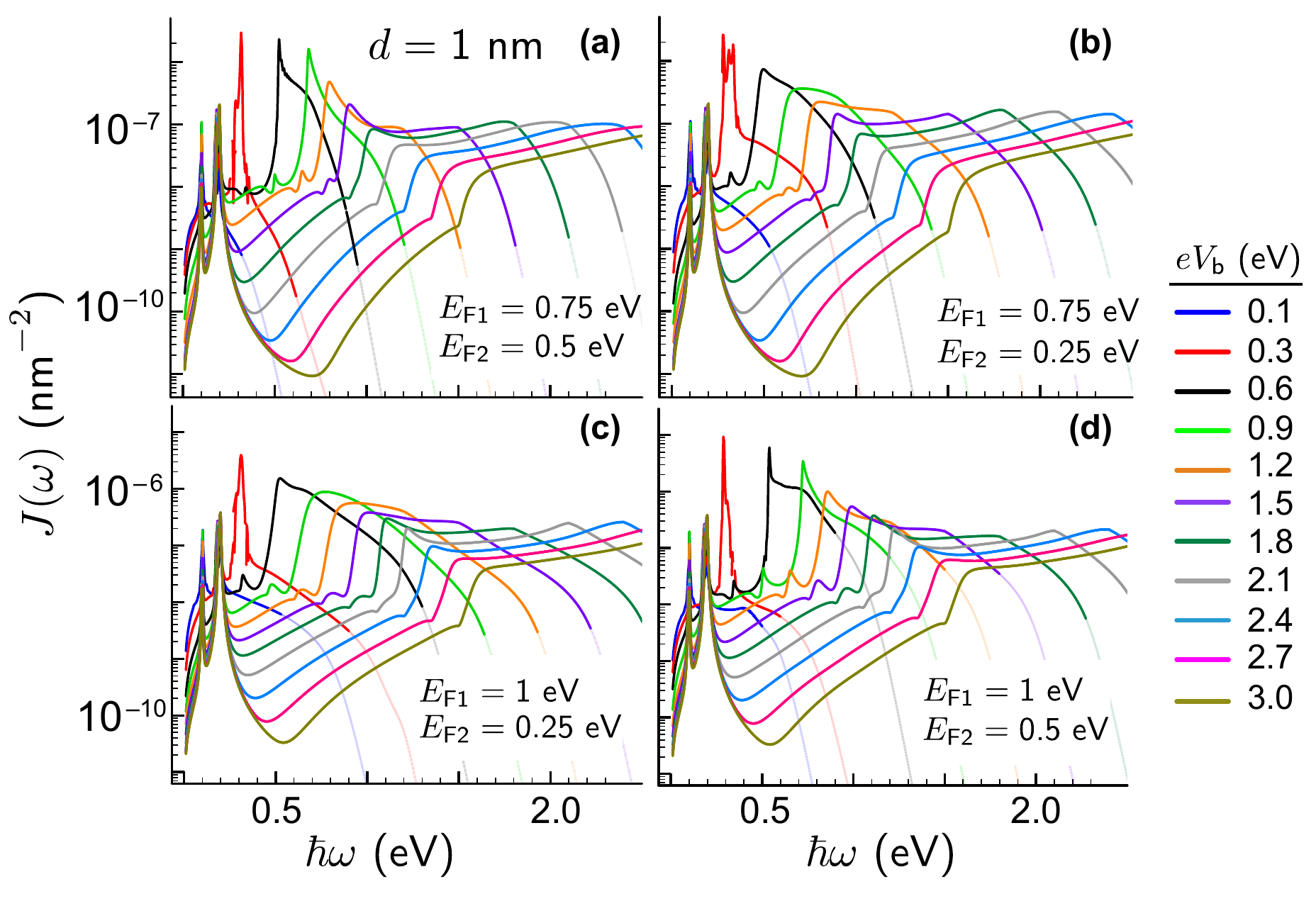}
\caption{{\bf Additional calculations of the spectrally resolved tunneling current.} Same as Fig.\ \ref{Fig4}a for fixed gap distance $d=1\,$nm and various combinations of the graphene Fermi energies (see labels).}
\label{FigS4}
\end{center}
\end{figure}

\begin{figure}
\begin{center}
\includegraphics[width=0.4\textwidth]{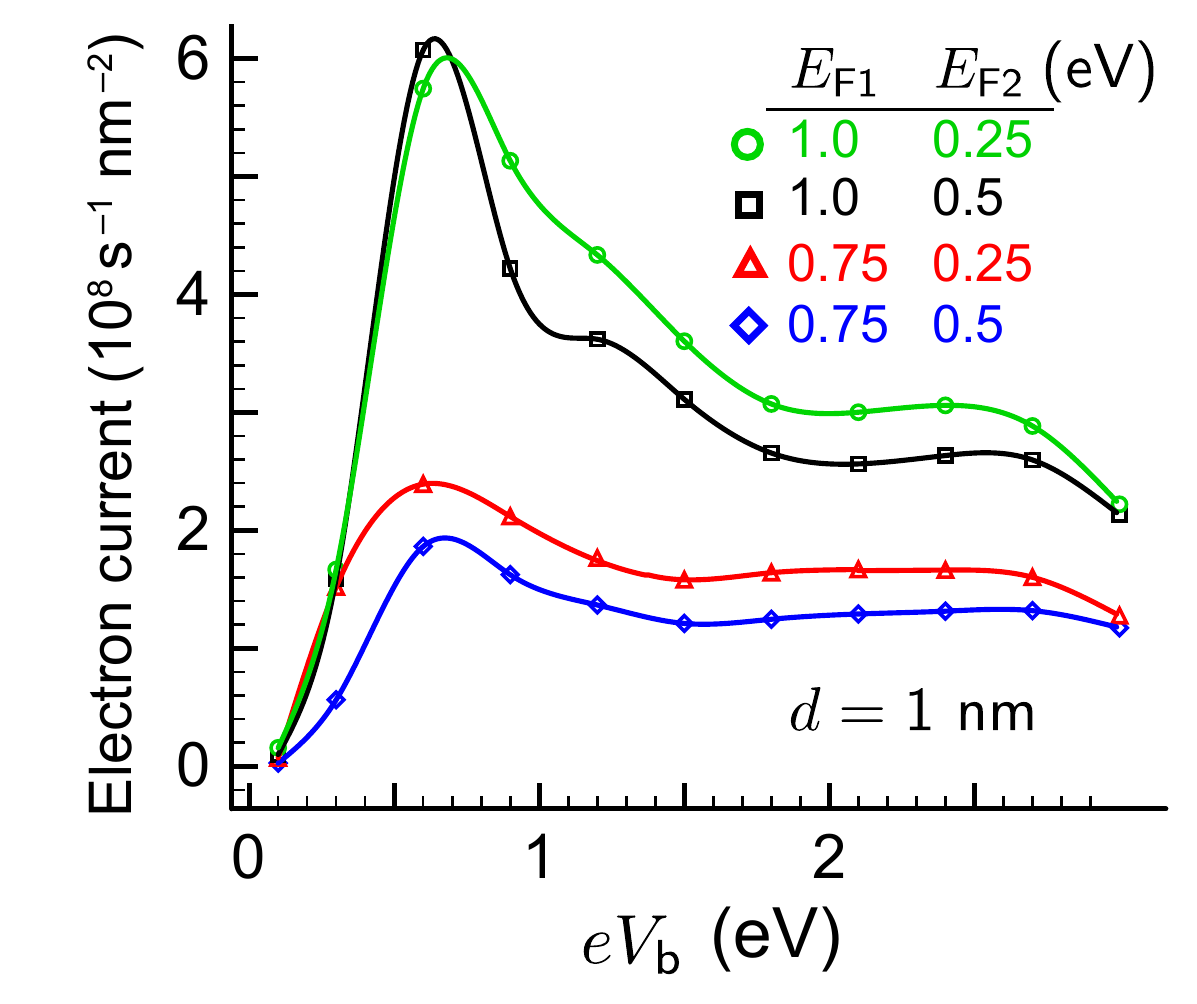}
\caption{{\bf Additional calculations of the tunneling current.} Same as Fig.\ \ref{Fig4}b for fixed gap distance $d=1\,$nm and various combinations of the graphene Fermi energies (see labels).}
\label{FigS5}
\end{center}
\end{figure}

\begin{figure*}
\begin{center}
\includegraphics[width=1\textwidth]{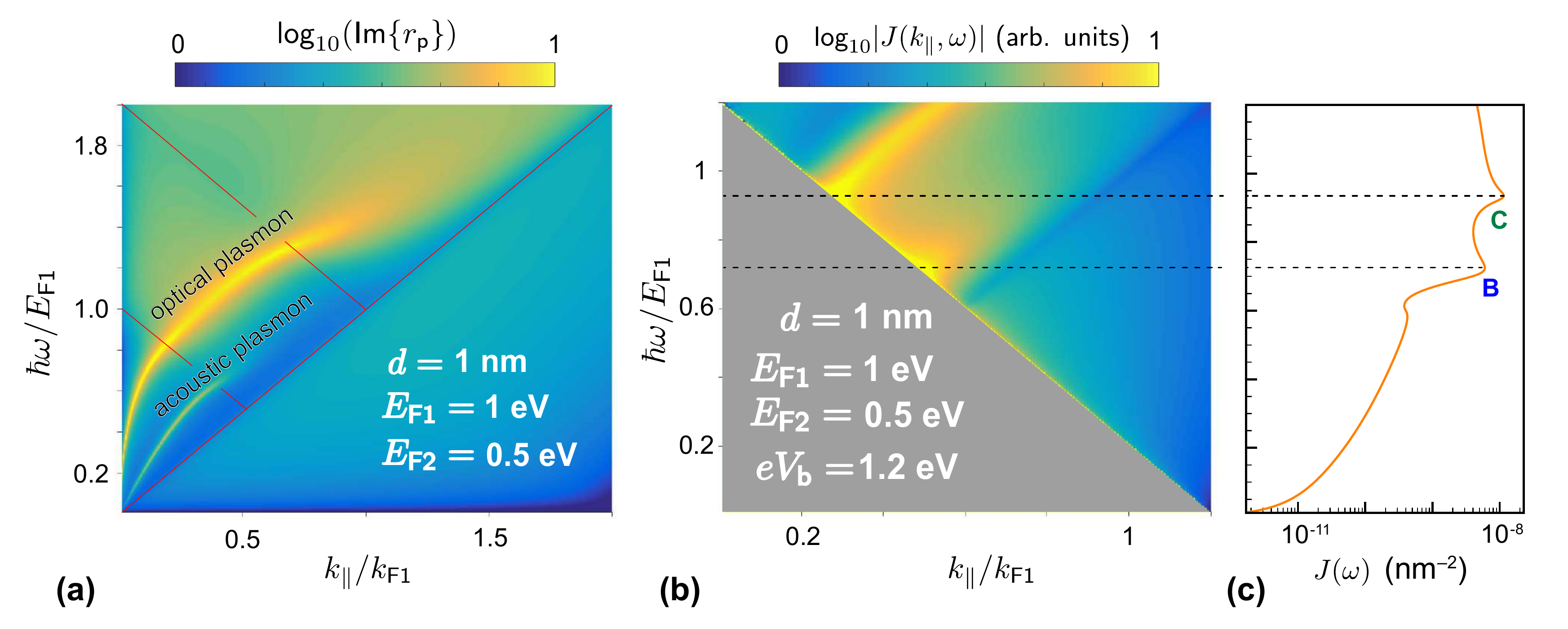}
\caption{{\bf Energy- and momentum-resolved electron tunneling with a vacuum gap.} Same as Fig.\ \ref{Fig2} with the hBN film replaced by vacuum.}
\label{FigS6}
\end{center}
\end{figure*}

\begin{figure*}
\begin{center}
\includegraphics[width=0.8\textwidth]{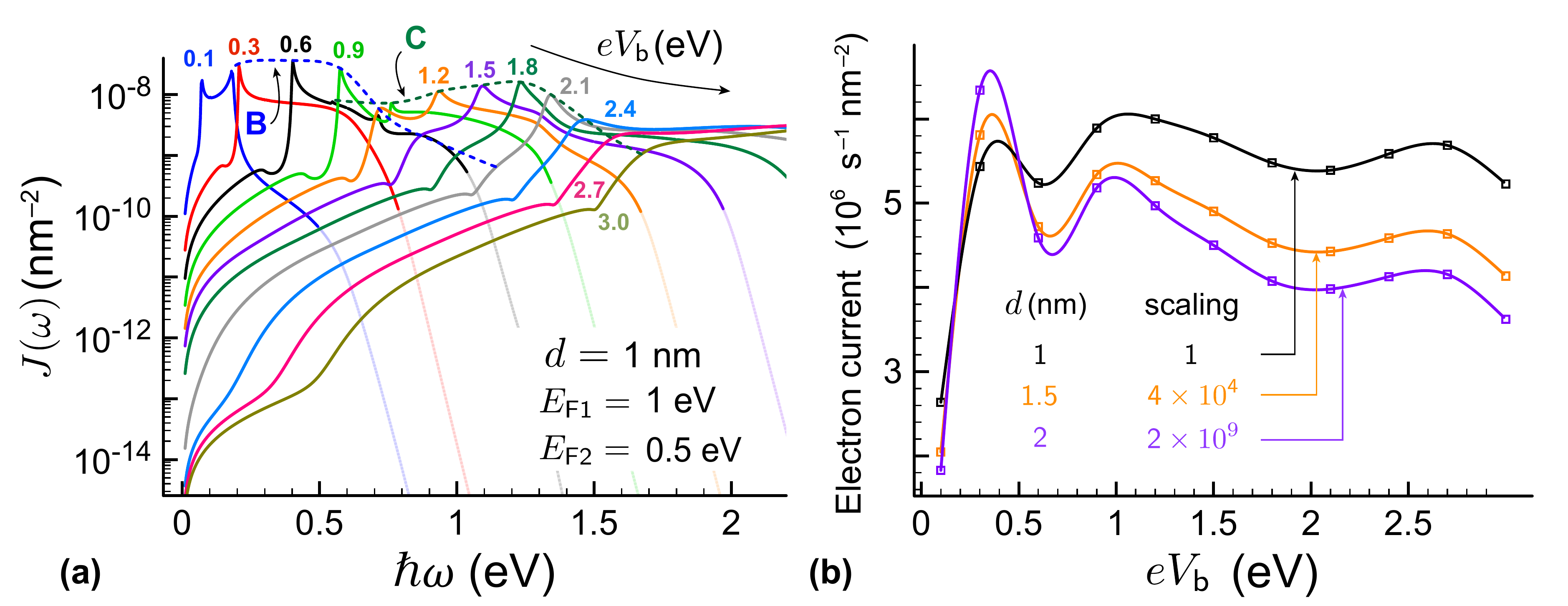}
\caption{{\bf Probability of plasmon-generation by electron tunneling with a vacuum gap.} Same as Fig.\ \ref{Fig4}a,b with the hBN film replaced by vacuum.}
\label{FigS7}
\end{center}
\end{figure*}

\section{Additional numerical simulations}

We present additional simulations of the dispersion diagrams and tunneling currents for various combinations of the graphene Fermi energies in Figs.\ \ref{FigS1}-\ref{FigS5}. We also show in Figs.\ \ref{FigS6} and \ref{FigS7} calculations similar to those of the Figs.\ \ref{Fig1}-\ref{Fig4} for graphene layers separated by vacuum instead of hBN. We assume a conservative graphene plasmon lifetime of $66\,$fs in all cases.

\acknowledgments

This work has been supported in part by the Spanish MINECO (MAT2014-59096-P and SEV2015-0522), the European Commission (Graphene Flagship CNECT-ICT-604391 and FP7-ICT-2013-613024-GRASP), Ag\`encia de Gesti\'o d'Ajuts Universitaris i de Recerca (AGAUR) (2014-SGR-1400), and Fundaci\'o Privada Cellex. SdV acknowledges financial support through the FPU program from the Spanish MECD.


\begin{thebibliography}{92}
\expandafter\ifx\csname natexlab\endcsname\relax\def\natexlab#1{#1}\fi
\expandafter\ifx\csname bibnamefont\endcsname\relax
  \def\bibnamefont#1{#1}\fi
\expandafter\ifx\csname bibfnamefont\endcsname\relax
  \def\bibfnamefont#1{#1}\fi
\expandafter\ifx\csname citenamefont\endcsname\relax
  \def\citenamefont#1{#1}\fi
\expandafter\ifx\csname url\endcsname\relax
  \def\url#1{\texttt{#1}}\fi
\expandafter\ifx\csname urlprefix\endcsname\relax\def\urlprefix{URL }\fi
\providecommand{\bibinfo}[2]{#2}
\providecommand{\eprint}[2][]{\url{#2}}

\bibitem[{\citenamefont{Li et~al.}(2003)\citenamefont{Li, Stockman, and
  Bergman}}]{LSB03}
\bibinfo{author}{\bibfnamefont{K.~R.} \bibnamefont{Li}},
  \bibinfo{author}{\bibfnamefont{M.~I.} \bibnamefont{Stockman}},
  \bibnamefont{and} \bibinfo{author}{\bibfnamefont{D.~J.}
  \bibnamefont{Bergman}}, \bibinfo{journal}{Phys.\ Rev.\ Lett.}
  \textbf{\bibinfo{volume}{91}}, \bibinfo{pages}{227402}
  (\bibinfo{year}{2003}).

\bibitem[{\citenamefont{{\'Alvarez-Puebla}
  et~al.}(2010)\citenamefont{{\'Alvarez-Puebla}, {Liz-Marz\'an}, and
  {Garc\'{\i}a de Abajo}}}]{paper156}
\bibinfo{author}{\bibfnamefont{R.~A.} \bibnamefont{{\'Alvarez-Puebla}}},
  \bibinfo{author}{\bibfnamefont{L.~M.} \bibnamefont{{Liz-Marz\'an}}},
  \bibnamefont{and} \bibinfo{author}{\bibfnamefont{F.~J.}
  \bibnamefont{{Garc\'{\i}a de Abajo}}}, \bibinfo{journal}{J.\ Phys.\ Chem.\
  Lett.} \textbf{\bibinfo{volume}{1}}, \bibinfo{pages}{2428}
  (\bibinfo{year}{2010}).

\bibitem[{\citenamefont{Liedberg et~al.}(1983)\citenamefont{Liedberg, Nylander,
  and Lunstr\"om}}]{LNL1983}
\bibinfo{author}{\bibfnamefont{B.}~\bibnamefont{Liedberg}},
  \bibinfo{author}{\bibfnamefont{C.}~\bibnamefont{Nylander}}, \bibnamefont{and}
  \bibinfo{author}{\bibfnamefont{I.}~\bibnamefont{Lunstr\"om}},
  \bibinfo{journal}{Sens.\ Actuators} \textbf{\bibinfo{volume}{4}},
  \bibinfo{pages}{299} (\bibinfo{year}{1983}).

\bibitem[{\citenamefont{Anker et~al.}(2008)\citenamefont{Anker, Hall, Lyandres,
  Shah, Zhao, and {Van Duyne}}}]{AHL08}
\bibinfo{author}{\bibfnamefont{J.~N.} \bibnamefont{Anker}},
  \bibinfo{author}{\bibfnamefont{W.~P.} \bibnamefont{Hall}},
  \bibinfo{author}{\bibfnamefont{O.}~\bibnamefont{Lyandres}},
  \bibinfo{author}{\bibfnamefont{N.~C.} \bibnamefont{Shah}},
  \bibinfo{author}{\bibfnamefont{J.}~\bibnamefont{Zhao}}, \bibnamefont{and}
  \bibinfo{author}{\bibfnamefont{R.~P.} \bibnamefont{{Van Duyne}}},
  \bibinfo{journal}{Nat.\ Mater.} \textbf{\bibinfo{volume}{7}},
  \bibinfo{pages}{442} (\bibinfo{year}{2008}).

\bibitem[{\citenamefont{Zeng et~al.}(2014)\citenamefont{Zeng, Baillargeat, Hod,
  and Yong}}]{ZBH14}
\bibinfo{author}{\bibfnamefont{S.}~\bibnamefont{Zeng}},
  \bibinfo{author}{\bibfnamefont{D.}~\bibnamefont{Baillargeat}},
  \bibinfo{author}{\bibfnamefont{H.-P.} \bibnamefont{Hod}}, \bibnamefont{and}
  \bibinfo{author}{\bibfnamefont{K.-T.} \bibnamefont{Yong}},
  \bibinfo{journal}{Chem.\ Soc.\ Rev.} \textbf{\bibinfo{volume}{43}},
  \bibinfo{pages}{3426} (\bibinfo{year}{2014}).

\bibitem[{\citenamefont{Reguera et~al.}(2017)\citenamefont{Reguera, {Jimenez de
  Aberasturi}, Langer, and Liz-Marz\'an}}]{RJL17}
\bibinfo{author}{\bibfnamefont{J.}~\bibnamefont{Reguera}},
  \bibinfo{author}{\bibfnamefont{D.}~\bibnamefont{{Jimenez de Aberasturi}}},
  \bibinfo{author}{\bibfnamefont{J.}~\bibnamefont{Langer}}, \bibnamefont{and}
  \bibinfo{author}{\bibfnamefont{L.~M.} \bibnamefont{Liz-Marz\'an}},
  \bibinfo{journal}{Chem.\ Soc.\ Rev.} \textbf{\bibinfo{volume}{in press}},
  \bibinfo{pages}{DOI: 10.1039/c7cs00158d} (\bibinfo{year}{2017}).

\bibitem[{\citenamefont{{O'Neal} et~al.}(2004)\citenamefont{{O'Neal}, Hirsch,
  Halas, Payne, and West}}]{NHH04}
\bibinfo{author}{\bibfnamefont{D.~P.} \bibnamefont{{O'Neal}}},
  \bibinfo{author}{\bibfnamefont{L.~R.} \bibnamefont{Hirsch}},
  \bibinfo{author}{\bibfnamefont{N.~J.} \bibnamefont{Halas}},
  \bibinfo{author}{\bibfnamefont{J.~D.} \bibnamefont{Payne}}, \bibnamefont{and}
  \bibinfo{author}{\bibfnamefont{J.~L.} \bibnamefont{West}},
  \bibinfo{journal}{Cancer\ Lett.} \textbf{\bibinfo{volume}{209}},
  \bibinfo{pages}{171} (\bibinfo{year}{2004}).

\bibitem[{\citenamefont{Jain et~al.}(2007)\citenamefont{Jain, El-Sayed, and
  El-Sayed}}]{JEE07}
\bibinfo{author}{\bibfnamefont{P.~K.} \bibnamefont{Jain}},
  \bibinfo{author}{\bibfnamefont{I.~H.} \bibnamefont{El-Sayed}},
  \bibnamefont{and} \bibinfo{author}{\bibfnamefont{M.~A.}
  \bibnamefont{El-Sayed}}, \bibinfo{journal}{Nano\ Today}
  \textbf{\bibinfo{volume}{2}}, \bibinfo{pages}{18} (\bibinfo{year}{2007}).

\bibitem[{\citenamefont{Luo et~al.}(2011)\citenamefont{Luo, Shiao, and
  Huang}}]{LSH11}
\bibinfo{author}{\bibfnamefont{Y.~L.} \bibnamefont{Luo}},
  \bibinfo{author}{\bibfnamefont{Y.~S.} \bibnamefont{Shiao}}, \bibnamefont{and}
  \bibinfo{author}{\bibfnamefont{Y.~F.} \bibnamefont{Huang}},
  \bibinfo{journal}{ACS\ Nano} \textbf{\bibinfo{volume}{5}},
  \bibinfo{pages}{7796} (\bibinfo{year}{2011}).

\bibitem[{\citenamefont{Palomba and Novotny}(2008)}]{PN08}
\bibinfo{author}{\bibfnamefont{S.}~\bibnamefont{Palomba}} \bibnamefont{and}
  \bibinfo{author}{\bibfnamefont{L.}~\bibnamefont{Novotny}},
  \bibinfo{journal}{Phys.\ Rev.\ Lett.} \textbf{\bibinfo{volume}{101}},
  \bibinfo{pages}{056802} (\bibinfo{year}{2008}).

\bibitem[{\citenamefont{Palomba et~al.}(2009)\citenamefont{Palomba, Danckwerts,
  and Novotny}}]{PDN09}
\bibinfo{author}{\bibfnamefont{S.}~\bibnamefont{Palomba}},
  \bibinfo{author}{\bibfnamefont{M.}~\bibnamefont{Danckwerts}},
  \bibnamefont{and} \bibinfo{author}{\bibfnamefont{L.}~\bibnamefont{Novotny}},
  \bibinfo{journal}{J.\ Opt.\ A:\ Pure\ Appl.\ Opt.}
  \textbf{\bibinfo{volume}{11}}, \bibinfo{pages}{114030}
  (\bibinfo{year}{2009}).

\bibitem[{\citenamefont{Shen et~al.}(2012)\citenamefont{Shen, Zhu, Yang, and
  Li}}]{SZY12}
\bibinfo{author}{\bibfnamefont{J.}~\bibnamefont{Shen}},
  \bibinfo{author}{\bibfnamefont{Y.}~\bibnamefont{Zhu}},
  \bibinfo{author}{\bibfnamefont{X.}~\bibnamefont{Yang}}, \bibnamefont{and}
  \bibinfo{author}{\bibfnamefont{C.}~\bibnamefont{Li}},
  \bibinfo{journal}{Chem.\ Commun.} \textbf{\bibinfo{volume}{48}},
  \bibinfo{pages}{3686} (\bibinfo{year}{2012}).

\bibitem[{\citenamefont{Wang et~al.}(2013)\citenamefont{Wang, Ranasingha,
  Natesakhawat, Ohodnicki, Andio, Lewis, and Matranga}}]{WRN13}
\bibinfo{author}{\bibfnamefont{C.}~\bibnamefont{Wang}},
  \bibinfo{author}{\bibfnamefont{O.}~\bibnamefont{Ranasingha}},
  \bibinfo{author}{\bibfnamefont{S.}~\bibnamefont{Natesakhawat}},
  \bibinfo{author}{\bibfnamefont{P.~R.} \bibnamefont{Ohodnicki}},
  \bibinfo{author}{\bibfnamefont{M.}~\bibnamefont{Andio}},
  \bibinfo{author}{\bibfnamefont{J.~P.} \bibnamefont{Lewis}}, \bibnamefont{and}
  \bibinfo{author}{\bibfnamefont{C.}~\bibnamefont{Matranga}},
  \bibinfo{journal}{Nanoscale} \textbf{\bibinfo{volume}{5}},
  \bibinfo{pages}{6968} (\bibinfo{year}{2013}).

\bibitem[{\citenamefont{Clavero}(2014)}]{C14}
\bibinfo{author}{\bibfnamefont{C.}~\bibnamefont{Clavero}},
  \bibinfo{journal}{Nat.\ Photon.} \textbf{\bibinfo{volume}{8}},
  \bibinfo{pages}{95} (\bibinfo{year}{2014}).

\bibitem[{\citenamefont{Park et~al.}(2015)\citenamefont{Park, Baker, and
  Somorjai}}]{PBS15}
\bibinfo{author}{\bibfnamefont{J.~Y.} \bibnamefont{Park}},
  \bibinfo{author}{\bibfnamefont{L.~R.} \bibnamefont{Baker}}, \bibnamefont{and}
  \bibinfo{author}{\bibfnamefont{G.~A.} \bibnamefont{Somorjai}},
  \bibinfo{journal}{Chem.\ Rev.} \textbf{\bibinfo{volume}{115}},
  \bibinfo{pages}{2781} (\bibinfo{year}{2015}).

\bibitem[{\citenamefont{Catchpole and Polman}(2008)}]{CP08}
\bibinfo{author}{\bibfnamefont{K.~R.} \bibnamefont{Catchpole}}
  \bibnamefont{and} \bibinfo{author}{\bibfnamefont{A.}~\bibnamefont{Polman}},
  \bibinfo{journal}{Opt.\ Express} \textbf{\bibinfo{volume}{16}},
  \bibinfo{pages}{21793} (\bibinfo{year}{2008}).

\bibitem[{\citenamefont{Atwater and Polman}(2010)}]{AP10}
\bibinfo{author}{\bibfnamefont{H.~A.} \bibnamefont{Atwater}} \bibnamefont{and}
  \bibinfo{author}{\bibfnamefont{A.}~\bibnamefont{Polman}},
  \bibinfo{journal}{Nat.\ Mater.} \textbf{\bibinfo{volume}{9}},
  \bibinfo{pages}{205} (\bibinfo{year}{2010}).

\bibitem[{\citenamefont{Fei et~al.}(2011)\citenamefont{Fei, Andreev, Bao,
  Zhang, McLeod, Wang, Stewart, Zhao, Dominguez, Thiemens et~al.}}]{FAB11}
\bibinfo{author}{\bibfnamefont{Z.}~\bibnamefont{Fei}},
  \bibinfo{author}{\bibfnamefont{G.~O.} \bibnamefont{Andreev}},
  \bibinfo{author}{\bibfnamefont{W.}~\bibnamefont{Bao}},
  \bibinfo{author}{\bibfnamefont{L.~M.} \bibnamefont{Zhang}},
  \bibinfo{author}{\bibfnamefont{A.~S.} \bibnamefont{McLeod}},
  \bibinfo{author}{\bibfnamefont{C.}~\bibnamefont{Wang}},
  \bibinfo{author}{\bibfnamefont{M.~K.} \bibnamefont{Stewart}},
  \bibinfo{author}{\bibfnamefont{Z.}~\bibnamefont{Zhao}},
  \bibinfo{author}{\bibfnamefont{G.}~\bibnamefont{Dominguez}},
  \bibinfo{author}{\bibfnamefont{M.}~\bibnamefont{Thiemens}},
  \bibnamefont{et~al.}, \bibinfo{journal}{Nano\ Lett.}
  \textbf{\bibinfo{volume}{11}}, \bibinfo{pages}{4701} (\bibinfo{year}{2011}).

\bibitem[{\citenamefont{Fei et~al.}(2012)\citenamefont{Fei, Rodin, Andreev,
  Bao, McLeod, Wagner, Zhang, Zhao, Thiemens, Dominguez et~al.}}]{FRA12}
\bibinfo{author}{\bibfnamefont{Z.}~\bibnamefont{Fei}},
  \bibinfo{author}{\bibfnamefont{A.~S.} \bibnamefont{Rodin}},
  \bibinfo{author}{\bibfnamefont{G.~O.} \bibnamefont{Andreev}},
  \bibinfo{author}{\bibfnamefont{W.}~\bibnamefont{Bao}},
  \bibinfo{author}{\bibfnamefont{A.~S.} \bibnamefont{McLeod}},
  \bibinfo{author}{\bibfnamefont{M.}~\bibnamefont{Wagner}},
  \bibinfo{author}{\bibfnamefont{L.~M.} \bibnamefont{Zhang}},
  \bibinfo{author}{\bibfnamefont{Z.}~\bibnamefont{Zhao}},
  \bibinfo{author}{\bibfnamefont{M.}~\bibnamefont{Thiemens}},
  \bibinfo{author}{\bibfnamefont{G.}~\bibnamefont{Dominguez}},
  \bibnamefont{et~al.}, \bibinfo{journal}{Nature}
  \textbf{\bibinfo{volume}{487}}, \bibinfo{pages}{82} (\bibinfo{year}{2012}).

\bibitem[{\citenamefont{Chen et~al.}(2012)\citenamefont{Chen, Badioli,
  Alonso-Gonz\'alez, Thongrattanasiri, Huth, Osmond, Spasenovi\'c, Centeno,
  Pesquera, Godignon et~al.}}]{paper196}
\bibinfo{author}{\bibfnamefont{J.}~\bibnamefont{Chen}},
  \bibinfo{author}{\bibfnamefont{M.}~\bibnamefont{Badioli}},
  \bibinfo{author}{\bibfnamefont{P.}~\bibnamefont{Alonso-Gonz\'alez}},
  \bibinfo{author}{\bibfnamefont{S.}~\bibnamefont{Thongrattanasiri}},
  \bibinfo{author}{\bibfnamefont{F.}~\bibnamefont{Huth}},
  \bibinfo{author}{\bibfnamefont{J.}~\bibnamefont{Osmond}},
  \bibinfo{author}{\bibfnamefont{M.}~\bibnamefont{Spasenovi\'c}},
  \bibinfo{author}{\bibfnamefont{A.}~\bibnamefont{Centeno}},
  \bibinfo{author}{\bibfnamefont{A.}~\bibnamefont{Pesquera}},
  \bibinfo{author}{\bibfnamefont{P.}~\bibnamefont{Godignon}},
  \bibnamefont{et~al.}, \bibinfo{journal}{Nature}
  \textbf{\bibinfo{volume}{487}}, \bibinfo{pages}{77} (\bibinfo{year}{2012}).

\bibitem[{\citenamefont{Fang et~al.}(2013)\citenamefont{Fang, Thongrattanasiri,
  Schlather, Liu, Ma, Wang, Ajayan, Nordlander, Halas, and {Garc\'{\i}a de
  Abajo}}}]{paper212}
\bibinfo{author}{\bibfnamefont{Z.}~\bibnamefont{Fang}},
  \bibinfo{author}{\bibfnamefont{S.}~\bibnamefont{Thongrattanasiri}},
  \bibinfo{author}{\bibfnamefont{A.}~\bibnamefont{Schlather}},
  \bibinfo{author}{\bibfnamefont{Z.}~\bibnamefont{Liu}},
  \bibinfo{author}{\bibfnamefont{L.}~\bibnamefont{Ma}},
  \bibinfo{author}{\bibfnamefont{Y.}~\bibnamefont{Wang}},
  \bibinfo{author}{\bibfnamefont{P.~M.} \bibnamefont{Ajayan}},
  \bibinfo{author}{\bibfnamefont{P.}~\bibnamefont{Nordlander}},
  \bibinfo{author}{\bibfnamefont{N.~J.} \bibnamefont{Halas}}, \bibnamefont{and}
  \bibinfo{author}{\bibfnamefont{F.~J.} \bibnamefont{{Garc\'{\i}a de Abajo}}},
  \bibinfo{journal}{ACS\ Nano} \textbf{\bibinfo{volume}{7}},
  \bibinfo{pages}{2388} (\bibinfo{year}{2013}).

\bibitem[{\citenamefont{Brar et~al.}(2013)\citenamefont{Brar, Jang, Sherrott,
  Lopez, and Atwater}}]{BJS13}
\bibinfo{author}{\bibfnamefont{V.~W.} \bibnamefont{Brar}},
  \bibinfo{author}{\bibfnamefont{M.~S.} \bibnamefont{Jang}},
  \bibinfo{author}{\bibfnamefont{M.}~\bibnamefont{Sherrott}},
  \bibinfo{author}{\bibfnamefont{J.~J.} \bibnamefont{Lopez}}, \bibnamefont{and}
  \bibinfo{author}{\bibfnamefont{H.~A.} \bibnamefont{Atwater}},
  \bibinfo{journal}{Nano\ Lett.} \textbf{\bibinfo{volume}{13}},
  \bibinfo{pages}{2541} (\bibinfo{year}{2013}).

\bibitem[{\citenamefont{Brar et~al.}(2014)\citenamefont{Brar, {Seok Jang},
  Sherrott, Kim, Lopez, Kim, Choi, and Atwater}}]{BSS14}
\bibinfo{author}{\bibfnamefont{V.~W.} \bibnamefont{Brar}},
  \bibinfo{author}{\bibfnamefont{M.}~\bibnamefont{{Seok Jang}}},
  \bibinfo{author}{\bibfnamefont{M.~C.} \bibnamefont{Sherrott}},
  \bibinfo{author}{\bibfnamefont{S.}~\bibnamefont{Kim}},
  \bibinfo{author}{\bibfnamefont{J.~J.} \bibnamefont{Lopez}},
  \bibinfo{author}{\bibfnamefont{L.~B.} \bibnamefont{Kim}},
  \bibinfo{author}{\bibfnamefont{M.}~\bibnamefont{Choi}}, \bibnamefont{and}
  \bibinfo{author}{\bibfnamefont{H.~A.} \bibnamefont{Atwater}},
  \bibinfo{journal}{Nano\ Lett.} \textbf{\bibinfo{volume}{14}},
  \bibinfo{pages}{3876} (\bibinfo{year}{2014}).

\bibitem[{\citenamefont{Yan et~al.}(2012)\citenamefont{Yan, Li, Li, Zhu,
  Avouris, and Xia}}]{YLL12}
\bibinfo{author}{\bibfnamefont{H.}~\bibnamefont{Yan}},
  \bibinfo{author}{\bibfnamefont{Z.}~\bibnamefont{Li}},
  \bibinfo{author}{\bibfnamefont{X.}~\bibnamefont{Li}},
  \bibinfo{author}{\bibfnamefont{W.}~\bibnamefont{Zhu}},
  \bibinfo{author}{\bibfnamefont{P.}~\bibnamefont{Avouris}}, \bibnamefont{and}
  \bibinfo{author}{\bibfnamefont{F.}~\bibnamefont{Xia}},
  \bibinfo{journal}{Nano\ Lett.} \textbf{\bibinfo{volume}{12}},
  \bibinfo{pages}{3766} (\bibinfo{year}{2012}).

\bibitem[{\citenamefont{Kumada et~al.}(2014)\citenamefont{Kumada, Roulleau,
  Roche, Hashisaka, Hibino, Petkovi\'{c}, and Glattli}}]{KRR14}
\bibinfo{author}{\bibfnamefont{N.}~\bibnamefont{Kumada}},
  \bibinfo{author}{\bibfnamefont{P.}~\bibnamefont{Roulleau}},
  \bibinfo{author}{\bibfnamefont{B.}~\bibnamefont{Roche}},
  \bibinfo{author}{\bibfnamefont{M.}~\bibnamefont{Hashisaka}},
  \bibinfo{author}{\bibfnamefont{H.}~\bibnamefont{Hibino}},
  \bibinfo{author}{\bibfnamefont{I.}~\bibnamefont{Petkovi\'{c}}},
  \bibnamefont{and} \bibinfo{author}{\bibfnamefont{D.~C.}
  \bibnamefont{Glattli}}, \bibinfo{journal}{Phys.\ Rev.\ Lett.}
  \textbf{\bibinfo{volume}{103}}, \bibinfo{pages}{266601}
  (\bibinfo{year}{2014}).

\bibitem[{\citenamefont{{Garc\'{\i}a de Abajo}}(2014)}]{paper235}
\bibinfo{author}{\bibfnamefont{F.~J.} \bibnamefont{{Garc\'{\i}a de Abajo}}},
  \bibinfo{journal}{ACS\ Photon.} \textbf{\bibinfo{volume}{1}},
  \bibinfo{pages}{135} (\bibinfo{year}{2014}).

\bibitem[{\citenamefont{Ni et~al.}(2016)\citenamefont{Ni, Wang, Goldflam,
  Wagner, Fei, McLeod, Liu, Keilmann, {\"Ozyilmaz}, Neto et~al.}}]{NWG16}
\bibinfo{author}{\bibfnamefont{G.~X.} \bibnamefont{Ni}},
  \bibinfo{author}{\bibfnamefont{L.}~\bibnamefont{Wang}},
  \bibinfo{author}{\bibfnamefont{M.~D.} \bibnamefont{Goldflam}},
  \bibinfo{author}{\bibfnamefont{M.}~\bibnamefont{Wagner}},
  \bibinfo{author}{\bibfnamefont{Z.}~\bibnamefont{Fei}},
  \bibinfo{author}{\bibfnamefont{A.~S.} \bibnamefont{McLeod}},
  \bibinfo{author}{\bibfnamefont{M.~K.} \bibnamefont{Liu}},
  \bibinfo{author}{\bibfnamefont{F.}~\bibnamefont{Keilmann}},
  \bibinfo{author}{\bibfnamefont{B.}~\bibnamefont{{\"Ozyilmaz}}},
  \bibinfo{author}{\bibfnamefont{A.~H.~C.} \bibnamefont{Neto}},
  \bibnamefont{et~al.}, \bibinfo{journal}{Nat.\ Photon.}
  \textbf{\bibinfo{volume}{10}}, \bibinfo{pages}{244} (\bibinfo{year}{2016}).

\bibitem[{\citenamefont{Woessner et~al.}(2015)\citenamefont{Woessner,
  Lundeberg, Gao, Principi, Alonso-Gonz\'alez, Carrega, Watanabe, Taniguchi,
  Vignale, Polini et~al.}}]{WLG15}
\bibinfo{author}{\bibfnamefont{A.}~\bibnamefont{Woessner}},
  \bibinfo{author}{\bibfnamefont{M.~B.} \bibnamefont{Lundeberg}},
  \bibinfo{author}{\bibfnamefont{Y.}~\bibnamefont{Gao}},
  \bibinfo{author}{\bibfnamefont{A.}~\bibnamefont{Principi}},
  \bibinfo{author}{\bibfnamefont{P.}~\bibnamefont{Alonso-Gonz\'alez}},
  \bibinfo{author}{\bibfnamefont{M.}~\bibnamefont{Carrega}},
  \bibinfo{author}{\bibfnamefont{K.}~\bibnamefont{Watanabe}},
  \bibinfo{author}{\bibfnamefont{T.}~\bibnamefont{Taniguchi}},
  \bibinfo{author}{\bibfnamefont{G.}~\bibnamefont{Vignale}},
  \bibinfo{author}{\bibfnamefont{M.}~\bibnamefont{Polini}},
  \bibnamefont{et~al.}, \bibinfo{journal}{Nat.\ Mater.}
  \textbf{\bibinfo{volume}{14}}, \bibinfo{pages}{421} (\bibinfo{year}{2015}).

\bibitem[{\citenamefont{Basov et~al.}(2016)\citenamefont{Basov, Fogler, and
  {Garc\'{\i}a de Abajo}}}]{paper283}
\bibinfo{author}{\bibfnamefont{D.~N.} \bibnamefont{Basov}},
  \bibinfo{author}{\bibfnamefont{M.~M.} \bibnamefont{Fogler}},
  \bibnamefont{and} \bibinfo{author}{\bibfnamefont{F.~J.}
  \bibnamefont{{Garc\'{\i}a de Abajo}}}, \bibinfo{journal}{Science}
  \textbf{\bibinfo{volume}{354}}, \bibinfo{pages}{aag1992}
  (\bibinfo{year}{2016}).

\bibitem[{\citenamefont{Yan et~al.}(2013)\citenamefont{Yan, Low, Zhu, Wu,
  Freitag, Li, Guinea, Avouris, and Xia}}]{YLZ12}
\bibinfo{author}{\bibfnamefont{H.}~\bibnamefont{Yan}},
  \bibinfo{author}{\bibfnamefont{T.}~\bibnamefont{Low}},
  \bibinfo{author}{\bibfnamefont{W.}~\bibnamefont{Zhu}},
  \bibinfo{author}{\bibfnamefont{Y.}~\bibnamefont{Wu}},
  \bibinfo{author}{\bibfnamefont{M.}~\bibnamefont{Freitag}},
  \bibinfo{author}{\bibfnamefont{X.}~\bibnamefont{Li}},
  \bibinfo{author}{\bibfnamefont{F.}~\bibnamefont{Guinea}},
  \bibinfo{author}{\bibfnamefont{P.}~\bibnamefont{Avouris}}, \bibnamefont{and}
  \bibinfo{author}{\bibfnamefont{F.}~\bibnamefont{Xia}},
  \bibinfo{journal}{Nat.\ Photon.} \textbf{\bibinfo{volume}{7}},
  \bibinfo{pages}{394} (\bibinfo{year}{2013}).

\bibitem[{\citenamefont{Freitag et~al.}(2010)\citenamefont{Freitag, Chiu,
  Steiner, Perebeinos, and Avouris}}]{FCS10}
\bibinfo{author}{\bibfnamefont{M.}~\bibnamefont{Freitag}},
  \bibinfo{author}{\bibfnamefont{H.-Y.} \bibnamefont{Chiu}},
  \bibinfo{author}{\bibfnamefont{M.}~\bibnamefont{Steiner}},
  \bibinfo{author}{\bibfnamefont{V.}~\bibnamefont{Perebeinos}},
  \bibnamefont{and} \bibinfo{author}{\bibfnamefont{P.}~\bibnamefont{Avouris}},
  \bibinfo{journal}{Nat.\ Nanotech.} \textbf{\bibinfo{volume}{5}},
  \bibinfo{pages}{497} (\bibinfo{year}{2010}).

\bibitem[{\citenamefont{Ilic et~al.}(2012)\citenamefont{Ilic, Jablan,
  Joannopoulos, Celanovic, Buljan, and Solja{\v{c}}i{\'c}}}]{IJJ12}
\bibinfo{author}{\bibfnamefont{O.}~\bibnamefont{Ilic}},
  \bibinfo{author}{\bibfnamefont{M.}~\bibnamefont{Jablan}},
  \bibinfo{author}{\bibfnamefont{J.~D.} \bibnamefont{Joannopoulos}},
  \bibinfo{author}{\bibfnamefont{I.}~\bibnamefont{Celanovic}},
  \bibinfo{author}{\bibfnamefont{H.}~\bibnamefont{Buljan}}, \bibnamefont{and}
  \bibinfo{author}{\bibfnamefont{M.}~\bibnamefont{Solja{\v{c}}i{\'c}}},
  \bibinfo{journal}{Phys.\ Rev.\ B} \textbf{\bibinfo{volume}{85}},
  \bibinfo{pages}{155422} (\bibinfo{year}{2012}).

\bibitem[{\citenamefont{Manjavacas et~al.}(2014)\citenamefont{Manjavacas,
  Thongrattanasiri, Greffet, and {Garc\'{\i}a de Abajo}}}]{paper245}
\bibinfo{author}{\bibfnamefont{A.}~\bibnamefont{Manjavacas}},
  \bibinfo{author}{\bibfnamefont{S.}~\bibnamefont{Thongrattanasiri}},
  \bibinfo{author}{\bibfnamefont{J.-J.} \bibnamefont{Greffet}},
  \bibnamefont{and} \bibinfo{author}{\bibfnamefont{F.~J.}
  \bibnamefont{{Garc\'{\i}a de Abajo}}}, \bibinfo{journal}{Appl.\ Phys.\ Lett.}
  \textbf{\bibinfo{volume}{105}}, \bibinfo{pages}{211102}
  (\bibinfo{year}{2014}).

\bibitem[{\citenamefont{Brar et~al.}(2015)\citenamefont{Brar, Sherrott, Jang,
  Kim, Kim, Choi, Sweatlock, and Atwater}}]{BSJ15}
\bibinfo{author}{\bibfnamefont{V.~W.} \bibnamefont{Brar}},
  \bibinfo{author}{\bibfnamefont{M.~C.} \bibnamefont{Sherrott}},
  \bibinfo{author}{\bibfnamefont{M.~S.} \bibnamefont{Jang}},
  \bibinfo{author}{\bibfnamefont{S.}~\bibnamefont{Kim}},
  \bibinfo{author}{\bibfnamefont{L.}~\bibnamefont{Kim}},
  \bibinfo{author}{\bibfnamefont{M.}~\bibnamefont{Choi}},
  \bibinfo{author}{\bibfnamefont{L.~A.} \bibnamefont{Sweatlock}},
  \bibnamefont{and} \bibinfo{author}{\bibfnamefont{H.~A.}
  \bibnamefont{Atwater}}, \bibinfo{journal}{Nat.\ Commun.}
  \textbf{\bibinfo{volume}{6}}, \bibinfo{pages}{7032} (\bibinfo{year}{2015}).

\bibitem[{\citenamefont{Yu and {Garc\'{\i}a de Abajo}}(2016)}]{paper275}
\bibinfo{author}{\bibfnamefont{R.}~\bibnamefont{Yu}} \bibnamefont{and}
  \bibinfo{author}{\bibfnamefont{F.~J.} \bibnamefont{{Garc\'{\i}a de Abajo}}},
  \bibinfo{journal}{ACS\ Nano} \textbf{\bibinfo{volume}{10}},
  \bibinfo{pages}{8045} (\bibinfo{year}{2016}).

\bibitem[{\citenamefont{Lundeberg et~al.}(2016)\citenamefont{Lundeberg, Gao,
  Woessner, Tan, Alonso-Gonz\'alez, Watanabe, Taniguchi, Hone, Hillenbrand, and
  Koppens}}]{LGW16}
\bibinfo{author}{\bibfnamefont{M.~B.} \bibnamefont{Lundeberg}},
  \bibinfo{author}{\bibfnamefont{Y.}~\bibnamefont{Gao}},
  \bibinfo{author}{\bibfnamefont{A.}~\bibnamefont{Woessner}},
  \bibinfo{author}{\bibfnamefont{C.}~\bibnamefont{Tan}},
  \bibinfo{author}{\bibfnamefont{P.}~\bibnamefont{Alonso-Gonz\'alez}},
  \bibinfo{author}{\bibfnamefont{K.}~\bibnamefont{Watanabe}},
  \bibinfo{author}{\bibfnamefont{T.}~\bibnamefont{Taniguchi}},
  \bibinfo{author}{\bibfnamefont{J.}~\bibnamefont{Hone}},
  \bibinfo{author}{\bibfnamefont{R.}~\bibnamefont{Hillenbrand}},
  \bibnamefont{and} \bibinfo{author}{\bibfnamefont{F.~H.~L.}
  \bibnamefont{Koppens}}, \bibinfo{journal}{Nat.\ Mater.}
  \textbf{\bibinfo{volume}{16}}, \bibinfo{pages}{204} (\bibinfo{year}{2016}).

\bibitem[{\citenamefont{Li et~al.}(2014)\citenamefont{Li, Yan, Farmer, Meng,
  Zhu, Osgood, Heinz, and Avouris}}]{LYF14}
\bibinfo{author}{\bibfnamefont{Y.}~\bibnamefont{Li}},
  \bibinfo{author}{\bibfnamefont{H.}~\bibnamefont{Yan}},
  \bibinfo{author}{\bibfnamefont{D.~B.} \bibnamefont{Farmer}},
  \bibinfo{author}{\bibfnamefont{X.}~\bibnamefont{Meng}},
  \bibinfo{author}{\bibfnamefont{W.}~\bibnamefont{Zhu}},
  \bibinfo{author}{\bibfnamefont{R.~M.} \bibnamefont{Osgood}},
  \bibinfo{author}{\bibfnamefont{T.~F.} \bibnamefont{Heinz}}, \bibnamefont{and}
  \bibinfo{author}{\bibfnamefont{P.}~\bibnamefont{Avouris}},
  \bibinfo{journal}{Nano\ Lett.} \textbf{\bibinfo{volume}{14}},
  \bibinfo{pages}{1573} (\bibinfo{year}{2014}).

\bibitem[{\citenamefont{Marini et~al.}(2015)\citenamefont{Marini, Silveiro, and
  {Garc\'{\i}a de Abajo}}}]{paper255}
\bibinfo{author}{\bibfnamefont{A.}~\bibnamefont{Marini}},
  \bibinfo{author}{\bibfnamefont{I.}~\bibnamefont{Silveiro}}, \bibnamefont{and}
  \bibinfo{author}{\bibfnamefont{F.~J.} \bibnamefont{{Garc\'{\i}a de Abajo}}},
  \bibinfo{journal}{ACS\ Photon.} \textbf{\bibinfo{volume}{2}},
  \bibinfo{pages}{876} (\bibinfo{year}{2015}).

\bibitem[{\citenamefont{Rodrigo et~al.}(2015)\citenamefont{Rodrigo, Limaj,
  Janner, Etezadi, {Garc\'{\i}a de Abajo}, Pruneri, and Altug}}]{paper256}
\bibinfo{author}{\bibfnamefont{D.}~\bibnamefont{Rodrigo}},
  \bibinfo{author}{\bibfnamefont{O.}~\bibnamefont{Limaj}},
  \bibinfo{author}{\bibfnamefont{D.}~\bibnamefont{Janner}},
  \bibinfo{author}{\bibfnamefont{D.}~\bibnamefont{Etezadi}},
  \bibinfo{author}{\bibfnamefont{F.~J.} \bibnamefont{{Garc\'{\i}a de Abajo}}},
  \bibinfo{author}{\bibfnamefont{V.}~\bibnamefont{Pruneri}}, \bibnamefont{and}
  \bibinfo{author}{\bibfnamefont{H.}~\bibnamefont{Altug}},
  \bibinfo{journal}{Science} \textbf{\bibinfo{volume}{349}},
  \bibinfo{pages}{165} (\bibinfo{year}{2015}).

\bibitem[{\citenamefont{Farmer et~al.}(2016)\citenamefont{Farmer, Avouris, Li,
  Heinz, and Han}}]{FAL16}
\bibinfo{author}{\bibfnamefont{D.~B.} \bibnamefont{Farmer}},
  \bibinfo{author}{\bibfnamefont{P.}~\bibnamefont{Avouris}},
  \bibinfo{author}{\bibfnamefont{Y.}~\bibnamefont{Li}},
  \bibinfo{author}{\bibfnamefont{T.~F.} \bibnamefont{Heinz}}, \bibnamefont{and}
  \bibinfo{author}{\bibfnamefont{S.-J.} \bibnamefont{Han}},
  \bibinfo{journal}{ACS\ Photon.} \textbf{\bibinfo{volume}{3}},
  \bibinfo{pages}{553} (\bibinfo{year}{2016}).

\bibitem[{\citenamefont{Chen et~al.}(2011)\citenamefont{Chen, Park, Boudouris,
  Horng, Geng, Girit, Zettl, Crommie, Segalman, Louie et~al.}}]{CPB11}
\bibinfo{author}{\bibfnamefont{C.~F.} \bibnamefont{Chen}},
  \bibinfo{author}{\bibfnamefont{C.~H.} \bibnamefont{Park}},
  \bibinfo{author}{\bibfnamefont{B.~W.} \bibnamefont{Boudouris}},
  \bibinfo{author}{\bibfnamefont{J.}~\bibnamefont{Horng}},
  \bibinfo{author}{\bibfnamefont{B.}~\bibnamefont{Geng}},
  \bibinfo{author}{\bibfnamefont{C.}~\bibnamefont{Girit}},
  \bibinfo{author}{\bibfnamefont{A.}~\bibnamefont{Zettl}},
  \bibinfo{author}{\bibfnamefont{M.~F.} \bibnamefont{Crommie}},
  \bibinfo{author}{\bibfnamefont{R.~A.} \bibnamefont{Segalman}},
  \bibinfo{author}{\bibfnamefont{S.~G.} \bibnamefont{Louie}},
  \bibnamefont{et~al.}, \bibinfo{journal}{Nature}
  \textbf{\bibinfo{volume}{471}}, \bibinfo{pages}{617} (\bibinfo{year}{2011}).

\bibitem[{\citenamefont{Ye et~al.}(2013)\citenamefont{Ye, Xiang, Lin, Peng,
  Huang, Yan, Cook, Samuel, Hwang, Ruan et~al.}}]{YXL13}
\bibinfo{author}{\bibfnamefont{R.}~\bibnamefont{Ye}},
  \bibinfo{author}{\bibfnamefont{C.}~\bibnamefont{Xiang}},
  \bibinfo{author}{\bibfnamefont{J.}~\bibnamefont{Lin}},
  \bibinfo{author}{\bibfnamefont{Z.}~\bibnamefont{Peng}},
  \bibinfo{author}{\bibfnamefont{K.}~\bibnamefont{Huang}},
  \bibinfo{author}{\bibfnamefont{Z.}~\bibnamefont{Yan}},
  \bibinfo{author}{\bibfnamefont{N.~P.} \bibnamefont{Cook}},
  \bibinfo{author}{\bibfnamefont{E.~L.} \bibnamefont{Samuel}},
  \bibinfo{author}{\bibfnamefont{C.-C.} \bibnamefont{Hwang}},
  \bibinfo{author}{\bibfnamefont{G.}~\bibnamefont{Ruan}}, \bibnamefont{et~al.},
  \bibinfo{journal}{Nat.\ Commun.} \textbf{\bibinfo{volume}{4}},
  \bibinfo{pages}{2943} (\bibinfo{year}{2013}).

\bibitem[{\citenamefont{M\"ullen}(2014)}]{M14}
\bibinfo{author}{\bibfnamefont{K.}~\bibnamefont{M\"ullen}},
  \bibinfo{journal}{ACS\ Nano} \textbf{\bibinfo{volume}{8}},
  \bibinfo{pages}{6531} (\bibinfo{year}{2014}).

\bibitem[{\citenamefont{Wang et~al.}(2016)\citenamefont{Wang, Li, Almdal,
  Mortensen, Xiao, and Ndoni}}]{WLA16}
\bibinfo{author}{\bibfnamefont{Z.}~\bibnamefont{Wang}},
  \bibinfo{author}{\bibfnamefont{T.}~\bibnamefont{Li}},
  \bibinfo{author}{\bibfnamefont{K.}~\bibnamefont{Almdal}},
  \bibinfo{author}{\bibfnamefont{N.~A.} \bibnamefont{Mortensen}},
  \bibinfo{author}{\bibfnamefont{S.}~\bibnamefont{Xiao}}, \bibnamefont{and}
  \bibinfo{author}{\bibfnamefont{S.}~\bibnamefont{Ndoni}},
  \bibinfo{journal}{Opt.\ Lett.} \textbf{\bibinfo{volume}{41}},
  \bibinfo{pages}{5345} (\bibinfo{year}{2016}).

\bibitem[{\citenamefont{Manjavacas et~al.}(2013)\citenamefont{Manjavacas,
  Marchesin, Thongrattanasiri, Koval, Nordlander, S\'{a}nchez-Portal, and
  {Garc\'{\i}a de Abajo}}}]{paper215}
\bibinfo{author}{\bibfnamefont{A.}~\bibnamefont{Manjavacas}},
  \bibinfo{author}{\bibfnamefont{F.}~\bibnamefont{Marchesin}},
  \bibinfo{author}{\bibfnamefont{S.}~\bibnamefont{Thongrattanasiri}},
  \bibinfo{author}{\bibfnamefont{P.}~\bibnamefont{Koval}},
  \bibinfo{author}{\bibfnamefont{P.}~\bibnamefont{Nordlander}},
  \bibinfo{author}{\bibfnamefont{D.}~\bibnamefont{S\'{a}nchez-Portal}},
  \bibnamefont{and} \bibinfo{author}{\bibfnamefont{F.~J.}
  \bibnamefont{{Garc\'{\i}a de Abajo}}}, \bibinfo{journal}{ACS\ Nano}
  \textbf{\bibinfo{volume}{7}}, \bibinfo{pages}{3635} (\bibinfo{year}{2013}).

\bibitem[{\citenamefont{Lauchner et~al.}(2015)\citenamefont{Lauchner,
  Schlather, Manjavacas, Cui, McClain, Stec, {Garc\'{\i}a de Abajo},
  Nordlander, and Halas}}]{paper260}
\bibinfo{author}{\bibfnamefont{A.}~\bibnamefont{Lauchner}},
  \bibinfo{author}{\bibfnamefont{A.}~\bibnamefont{Schlather}},
  \bibinfo{author}{\bibfnamefont{A.}~\bibnamefont{Manjavacas}},
  \bibinfo{author}{\bibfnamefont{Y.}~\bibnamefont{Cui}},
  \bibinfo{author}{\bibfnamefont{M.~J.} \bibnamefont{McClain}},
  \bibinfo{author}{\bibfnamefont{G.~J.} \bibnamefont{Stec}},
  \bibinfo{author}{\bibfnamefont{F.~J.} \bibnamefont{{Garc\'{\i}a de Abajo}}},
  \bibinfo{author}{\bibfnamefont{P.}~\bibnamefont{Nordlander}},
  \bibnamefont{and} \bibinfo{author}{\bibfnamefont{N.~J.} \bibnamefont{Halas}},
  \bibinfo{journal}{Nano\ Lett.} \textbf{\bibinfo{volume}{15}},
  \bibinfo{pages}{6208} (\bibinfo{year}{2015}).

\bibitem[{\citenamefont{Fei et~al.}(2017)\citenamefont{Fei, Ni, Jiang, Fogler,
  and Basov}}]{FNJ17}
\bibinfo{author}{\bibfnamefont{Z.}~\bibnamefont{Fei}},
  \bibinfo{author}{\bibfnamefont{G.-X.} \bibnamefont{Ni}},
  \bibinfo{author}{\bibfnamefont{B.-Y.} \bibnamefont{Jiang}},
  \bibinfo{author}{\bibfnamefont{M.~M.} \bibnamefont{Fogler}},
  \bibnamefont{and} \bibinfo{author}{\bibfnamefont{D.~N.} \bibnamefont{Basov}},
  \bibinfo{journal}{ACS\ Photon.} p. \bibinfo{pages}{DOI:
  10.1021/acsphotonics.7b00477} (\bibinfo{year}{2017}).

\bibitem[{\citenamefont{{Garc\'{\i}a de Abajo}}(2010)}]{paper149}
\bibinfo{author}{\bibfnamefont{F.~J.} \bibnamefont{{Garc\'{\i}a de Abajo}}},
  \bibinfo{journal}{Rev.\ Mod.\ Phys.} \textbf{\bibinfo{volume}{82}},
  \bibinfo{pages}{209} (\bibinfo{year}{2010}).

\bibitem[{\citenamefont{Tsui}(1969)}]{T1969}
\bibinfo{author}{\bibfnamefont{D.~C.} \bibnamefont{Tsui}},
  \bibinfo{journal}{Phys.\ Rev.\ Lett.} \textbf{\bibinfo{volume}{22}},
  \bibinfo{pages}{293} (\bibinfo{year}{1969}).

\bibitem[{\citenamefont{Tsui and {Barker Jr}}(1969)}]{TB1969}
\bibinfo{author}{\bibfnamefont{D.~C.} \bibnamefont{Tsui}} \bibnamefont{and}
  \bibinfo{author}{\bibfnamefont{A.~S.} \bibnamefont{{Barker Jr}}},
  \bibinfo{journal}{Phys.\ Rev.} \textbf{\bibinfo{volume}{186}},
  \bibinfo{pages}{590} (\bibinfo{year}{1969}).

\bibitem[{\citenamefont{Duke}(1969)}]{D1969}
\bibinfo{author}{\bibfnamefont{C.~B.} \bibnamefont{Duke}},
  \bibinfo{journal}{Phys.\ Rev.} \textbf{\bibinfo{volume}{186}},
  \bibinfo{pages}{588} (\bibinfo{year}{1969}).

\bibitem[{\citenamefont{Duke et~al.}(1969)\citenamefont{Duke, Rice, and
  Steinrisser}}]{DRS1969}
\bibinfo{author}{\bibfnamefont{C.~B.} \bibnamefont{Duke}},
  \bibinfo{author}{\bibfnamefont{M.~J.} \bibnamefont{Rice}}, \bibnamefont{and}
  \bibinfo{author}{\bibfnamefont{F.}~\bibnamefont{Steinrisser}},
  \bibinfo{journal}{Phys.\ Rev.} \textbf{\bibinfo{volume}{181}},
  \bibinfo{pages}{733} (\bibinfo{year}{1969}).

\bibitem[{\citenamefont{Economou and Ngai}(1971)}]{EN1971}
\bibinfo{author}{\bibfnamefont{E.~N.} \bibnamefont{Economou}} \bibnamefont{and}
  \bibinfo{author}{\bibfnamefont{K.~L.} \bibnamefont{Ngai}},
  \bibinfo{journal}{Phys.\ Rev.\ B} \textbf{\bibinfo{volume}{4}},
  \bibinfo{pages}{4105} (\bibinfo{year}{1971}).

\bibitem[{\citenamefont{Johansson et~al.}(1990)\citenamefont{Johansson,
  Monreal, and Apell}}]{JMA1990}
\bibinfo{author}{\bibfnamefont{P.}~\bibnamefont{Johansson}},
  \bibinfo{author}{\bibfnamefont{R.}~\bibnamefont{Monreal}}, \bibnamefont{and}
  \bibinfo{author}{\bibfnamefont{P.}~\bibnamefont{Apell}},
  \bibinfo{journal}{Phys.\ Rev.\ Lett.} \textbf{\bibinfo{volume}{42}},
  \bibinfo{pages}{9210} (\bibinfo{year}{1990}).

\bibitem[{\citenamefont{Berndt et~al.}(1991)\citenamefont{Berndt, Gimzewski,
  and Johansson}}]{BGJ91}
\bibinfo{author}{\bibfnamefont{R.}~\bibnamefont{Berndt}},
  \bibinfo{author}{\bibfnamefont{J.~K.} \bibnamefont{Gimzewski}},
  \bibnamefont{and}
  \bibinfo{author}{\bibfnamefont{P.}~\bibnamefont{Johansson}},
  \bibinfo{journal}{Phys.\ Rev.\ Lett.} \textbf{\bibinfo{volume}{67}},
  \bibinfo{pages}{3796} (\bibinfo{year}{1991}).

\bibitem[{\citenamefont{Rai et~al.}(2013)\citenamefont{Rai, Hartmann,
  Berthelot, Arocas, {Colas des Francs}, Hartschuh, and Bouhelier}}]{RHB13}
\bibinfo{author}{\bibfnamefont{P.}~\bibnamefont{Rai}},
  \bibinfo{author}{\bibfnamefont{N.}~\bibnamefont{Hartmann}},
  \bibinfo{author}{\bibfnamefont{J.}~\bibnamefont{Berthelot}},
  \bibinfo{author}{\bibfnamefont{J.}~\bibnamefont{Arocas}},
  \bibinfo{author}{\bibfnamefont{G.}~\bibnamefont{{Colas des Francs}}},
  \bibinfo{author}{\bibfnamefont{A.}~\bibnamefont{Hartschuh}},
  \bibnamefont{and}
  \bibinfo{author}{\bibfnamefont{A.}~\bibnamefont{Bouhelier}},
  \bibinfo{journal}{Phys.\ Rev.\ Lett.} \textbf{\bibinfo{volume}{111}},
  \bibinfo{pages}{026804} (\bibinfo{year}{2013}).

\bibitem[{\citenamefont{Ooi et~al.}(2015)\citenamefont{Ooi, Chu, Hsieh, Tan,
  and Ang}}]{OCH15}
\bibinfo{author}{\bibfnamefont{K.~J.~A.} \bibnamefont{Ooi}},
  \bibinfo{author}{\bibfnamefont{H.~S.} \bibnamefont{Chu}},
  \bibinfo{author}{\bibfnamefont{C.~Y.} \bibnamefont{Hsieh}},
  \bibinfo{author}{\bibfnamefont{D.~T.~H.} \bibnamefont{Tan}},
  \bibnamefont{and} \bibinfo{author}{\bibfnamefont{L.~K.} \bibnamefont{Ang}},
  \bibinfo{journal}{Phys.\ Rev.\ Applied} \textbf{\bibinfo{volume}{3}},
  \bibinfo{pages}{054001} (\bibinfo{year}{2015}).

\bibitem[{\citenamefont{Schneider et~al.}(2010)\citenamefont{Schneider, Schull,
  and Berndt}}]{SSB10}
\bibinfo{author}{\bibfnamefont{N.~L.} \bibnamefont{Schneider}},
  \bibinfo{author}{\bibfnamefont{G.}~\bibnamefont{Schull}}, \bibnamefont{and}
  \bibinfo{author}{\bibfnamefont{R.}~\bibnamefont{Berndt}},
  \bibinfo{journal}{Phys.\ Rev.\ Lett.} \textbf{\bibinfo{volume}{105}},
  \bibinfo{pages}{026601} (\bibinfo{year}{2010}).

\bibitem[{\citenamefont{Bharadwaj et~al.}(2011)\citenamefont{Bharadwaj,
  Bouhelier, and Novotny}}]{BBN11}
\bibinfo{author}{\bibfnamefont{P.}~\bibnamefont{Bharadwaj}},
  \bibinfo{author}{\bibfnamefont{A.}~\bibnamefont{Bouhelier}},
  \bibnamefont{and} \bibinfo{author}{\bibfnamefont{L.}~\bibnamefont{Novotny}},
  \bibinfo{journal}{Phys.\ Rev.\ Lett.} \textbf{\bibinfo{volume}{106}},
  \bibinfo{pages}{226802} (\bibinfo{year}{2011}).

\bibitem[{\citenamefont{Gro$\beta$e et~al.}(2014)\citenamefont{Gro$\beta$e,
  Kabakchiev, Lutz, Froidevaux, Schramm, Ruben, Etzkorn, Schlickum, Kuhnke,
  et~al.}}]{GKL14}
\bibinfo{author}{\bibfnamefont{C.}~\bibnamefont{Gro$\beta$e}},
  \bibinfo{author}{\bibfnamefont{A.}~\bibnamefont{Kabakchiev}},
  \bibinfo{author}{\bibfnamefont{T.}~\bibnamefont{Lutz}},
  \bibinfo{author}{\bibfnamefont{R.}~\bibnamefont{Froidevaux}},
  \bibinfo{author}{\bibfnamefont{F.}~\bibnamefont{Schramm}},
  \bibinfo{author}{\bibfnamefont{M.}~\bibnamefont{Ruben}},
  \bibinfo{author}{\bibfnamefont{M.}~\bibnamefont{Etzkorn}},
  \bibinfo{author}{\bibfnamefont{U.}~\bibnamefont{Schlickum}},
  \bibinfo{author}{\bibfnamefont{K.}~\bibnamefont{Kuhnke}}, ,
  \bibnamefont{et~al.}, \bibinfo{journal}{Nano\ Lett.}
  \textbf{\bibinfo{volume}{14}}, \bibinfo{pages}{5693} (\bibinfo{year}{2014}).

\bibitem[{\citenamefont{Kern et~al.}(2015)\citenamefont{Kern, Kullock,
  Prangsma, Emmerling, Kamp, and Hecht}}]{KKP15}
\bibinfo{author}{\bibfnamefont{J.}~\bibnamefont{Kern}},
  \bibinfo{author}{\bibfnamefont{R.}~\bibnamefont{Kullock}},
  \bibinfo{author}{\bibfnamefont{J.}~\bibnamefont{Prangsma}},
  \bibinfo{author}{\bibfnamefont{M.}~\bibnamefont{Emmerling}},
  \bibinfo{author}{\bibfnamefont{M.}~\bibnamefont{Kamp}}, \bibnamefont{and}
  \bibinfo{author}{\bibfnamefont{B.}~\bibnamefont{Hecht}},
  \bibinfo{journal}{Nat.\ Photon.} \textbf{\bibinfo{volume}{9}},
  \bibinfo{pages}{582} (\bibinfo{year}{2015}).

\bibitem[{\citenamefont{Uskov et~al.}(2017)\citenamefont{Uskov, Khurgin, Buret,
  Bouhelier, Smetanin, and Protsenko}}]{UKB17}
\bibinfo{author}{\bibfnamefont{A.~V.} \bibnamefont{Uskov}},
  \bibinfo{author}{\bibfnamefont{J.~B.} \bibnamefont{Khurgin}},
  \bibinfo{author}{\bibfnamefont{M.}~\bibnamefont{Buret}},
  \bibinfo{author}{\bibfnamefont{A.}~\bibnamefont{Bouhelier}},
  \bibinfo{author}{\bibfnamefont{I.~V.} \bibnamefont{Smetanin}},
  \bibnamefont{and} \bibinfo{author}{\bibfnamefont{I.~E.}
  \bibnamefont{Protsenko}}, \bibinfo{journal}{ACS\ Photon.}
  \textbf{\bibinfo{volume}{4}}, \bibinfo{pages}{1501} (\bibinfo{year}{2017}).

\bibitem[{\citenamefont{Koller et~al.}(2008)\citenamefont{Koller, Hohenau,
  Ditlbacher, Galler, Reil, Aussenegg, Leitner, List, and Krenn}}]{KHD08}
\bibinfo{author}{\bibfnamefont{D.~M.} \bibnamefont{Koller}},
  \bibinfo{author}{\bibfnamefont{A.}~\bibnamefont{Hohenau}},
  \bibinfo{author}{\bibfnamefont{H.}~\bibnamefont{Ditlbacher}},
  \bibinfo{author}{\bibfnamefont{N.}~\bibnamefont{Galler}},
  \bibinfo{author}{\bibfnamefont{F.}~\bibnamefont{Reil}},
  \bibinfo{author}{\bibfnamefont{F.~R.} \bibnamefont{Aussenegg}},
  \bibinfo{author}{\bibfnamefont{A.}~\bibnamefont{Leitner}},
  \bibinfo{author}{\bibfnamefont{E.~J.~W.} \bibnamefont{List}},
  \bibnamefont{and} \bibinfo{author}{\bibfnamefont{J.~R.} \bibnamefont{Krenn}},
  \bibinfo{journal}{Nat.\ Photon.} \textbf{\bibinfo{volume}{2}},
  \bibinfo{pages}{684} (\bibinfo{year}{2008}).

\bibitem[{\citenamefont{Walters et~al.}(2010)\citenamefont{Walters, {van Loon},
  Brunets, Schmitz, and Polman}}]{WVB10}
\bibinfo{author}{\bibfnamefont{R.~J.} \bibnamefont{Walters}},
  \bibinfo{author}{\bibfnamefont{R.~V.~A.} \bibnamefont{{van Loon}}},
  \bibinfo{author}{\bibfnamefont{I.}~\bibnamefont{Brunets}},
  \bibinfo{author}{\bibfnamefont{J.}~\bibnamefont{Schmitz}}, \bibnamefont{and}
  \bibinfo{author}{\bibfnamefont{A.}~\bibnamefont{Polman}},
  \bibinfo{journal}{Nat.\ Mater.} \textbf{\bibinfo{volume}{9}},
  \bibinfo{pages}{21} (\bibinfo{year}{2010}).

\bibitem[{\citenamefont{Enaldiev et~al.}(2017)\citenamefont{Enaldiev, Bylinkin,
  and Svintsov}}]{EBS17}
\bibinfo{author}{\bibfnamefont{V.}~\bibnamefont{Enaldiev}},
  \bibinfo{author}{\bibfnamefont{A.}~\bibnamefont{Bylinkin}}, \bibnamefont{and}
  \bibinfo{author}{\bibfnamefont{D.}~\bibnamefont{Svintsov}},
  \bibinfo{journal}{0} \textbf{\bibinfo{volume}{0}},
  \bibinfo{pages}{arXiv:1706.05216v1} (\bibinfo{year}{2017}).

\bibitem[{\citenamefont{Svintsov et~al.}(2016)\citenamefont{Svintsov,
  Devizorova, Otsuji, and Ryzhii}}]{SDR16}
\bibinfo{author}{\bibfnamefont{D.}~\bibnamefont{Svintsov}},
  \bibinfo{author}{\bibfnamefont{Z.}~\bibnamefont{Devizorova}},
  \bibinfo{author}{\bibfnamefont{T.}~\bibnamefont{Otsuji}}, \bibnamefont{and}
  \bibinfo{author}{\bibfnamefont{V.}~\bibnamefont{Ryzhii}},
  \bibinfo{journal}{Phys.\ Rev.\ B} \textbf{\bibinfo{volume}{94}},
  \bibinfo{pages}{115301} (\bibinfo{year}{2016}).

\bibitem[{\citenamefont{Yadav et~al.}(2016)\citenamefont{Yadav, Tombet,
  Watanabe, Arnold, Ryzhii, and Otsuji}}]{YTW16}
\bibinfo{author}{\bibfnamefont{D.}~\bibnamefont{Yadav}},
  \bibinfo{author}{\bibfnamefont{S.~B.} \bibnamefont{Tombet}},
  \bibinfo{author}{\bibfnamefont{T.}~\bibnamefont{Watanabe}},
  \bibinfo{author}{\bibfnamefont{S.}~\bibnamefont{Arnold}},
  \bibinfo{author}{\bibfnamefont{V.}~\bibnamefont{Ryzhii}}, \bibnamefont{and}
  \bibinfo{author}{\bibfnamefont{T.}~\bibnamefont{Otsuji}},
  \bibinfo{journal}{2D\ Mater.} \textbf{\bibinfo{volume}{3}},
  \bibinfo{pages}{045009} (\bibinfo{year}{2016}).

\bibitem[{\citenamefont{Wang et~al.}(2015)\citenamefont{Wang, Braun, Zhang,
  Peisert, Adler, Chass\'e, and Meixner}}]{WBZ15}
\bibinfo{author}{\bibfnamefont{X.}~\bibnamefont{Wang}},
  \bibinfo{author}{\bibfnamefont{K.}~\bibnamefont{Braun}},
  \bibinfo{author}{\bibfnamefont{D.}~\bibnamefont{Zhang}},
  \bibinfo{author}{\bibfnamefont{H.}~\bibnamefont{Peisert}},
  \bibinfo{author}{\bibfnamefont{H.}~\bibnamefont{Adler}},
  \bibinfo{author}{\bibfnamefont{T.}~\bibnamefont{Chass\'e}}, \bibnamefont{and}
  \bibinfo{author}{\bibfnamefont{A.~J.} \bibnamefont{Meixner}},
  \bibinfo{journal}{ACS\ Nano} \textbf{\bibinfo{volume}{9}},
  \bibinfo{pages}{8176} (\bibinfo{year}{2015}).

\bibitem[{\citenamefont{Aussenegg et~al.}(2006)\citenamefont{Aussenegg, Krenn,
  Jakopic, and Leising}}]{DAK06}
\bibinfo{author}{\bibfnamefont{H.~D. F.~R.} \bibnamefont{Aussenegg}},
  \bibinfo{author}{\bibfnamefont{J.~R.} \bibnamefont{Krenn}},
  \bibinfo{author}{\bibfnamefont{L.~G.} \bibnamefont{Jakopic}},
  \bibnamefont{and} \bibinfo{author}{\bibfnamefont{G.}~\bibnamefont{Leising}},
  \bibinfo{journal}{Appl.\ Phys.\ Lett.} \textbf{\bibinfo{volume}{89}},
  \bibinfo{pages}{161101} (\bibinfo{year}{2006}).

\bibitem[{\citenamefont{Heeres et~al.}(2009)\citenamefont{Heeres, Dorenbos,
  Koene, Solomon, Kouwenhoven, and Zwiller}}]{HDK09_2}
\bibinfo{author}{\bibfnamefont{R.~W.} \bibnamefont{Heeres}},
  \bibinfo{author}{\bibfnamefont{S.~N.} \bibnamefont{Dorenbos}},
  \bibinfo{author}{\bibfnamefont{B.}~\bibnamefont{Koene}},
  \bibinfo{author}{\bibfnamefont{G.~S.} \bibnamefont{Solomon}},
  \bibinfo{author}{\bibfnamefont{L.~P.} \bibnamefont{Kouwenhoven}},
  \bibnamefont{and} \bibinfo{author}{\bibfnamefont{V.}~\bibnamefont{Zwiller}},
  \bibinfo{journal}{Nano\ Lett.} \textbf{\bibinfo{volume}{10}},
  \bibinfo{pages}{661} (\bibinfo{year}{2009}).

\bibitem[{\citenamefont{Neutens et~al.}(2009)\citenamefont{Neutens, {Van
  Dorpe}, {De Vlaminck}, Lagae, and Borghs}}]{NVD09}
\bibinfo{author}{\bibfnamefont{P.}~\bibnamefont{Neutens}},
  \bibinfo{author}{\bibfnamefont{P.}~\bibnamefont{{Van Dorpe}}},
  \bibinfo{author}{\bibfnamefont{I.}~\bibnamefont{{De Vlaminck}}},
  \bibinfo{author}{\bibfnamefont{L.}~\bibnamefont{Lagae}}, \bibnamefont{and}
  \bibinfo{author}{\bibfnamefont{G.}~\bibnamefont{Borghs}},
  \bibinfo{journal}{Nat.\ Photon.} \textbf{\bibinfo{volume}{3}},
  \bibinfo{pages}{283} (\bibinfo{year}{2009}).

\bibitem[{\citenamefont{Falk et~al.}(2009)\citenamefont{Falk, Koppens, Chun,
  Kang, de~Leon~Snapp, Akimov, Jo, Lukin, and Park}}]{FKC09}
\bibinfo{author}{\bibfnamefont{A.~L.} \bibnamefont{Falk}},
  \bibinfo{author}{\bibfnamefont{F.~H.} \bibnamefont{Koppens}},
  \bibinfo{author}{\bibfnamefont{L.~Y.} \bibnamefont{Chun}},
  \bibinfo{author}{\bibfnamefont{K.}~\bibnamefont{Kang}},
  \bibinfo{author}{\bibfnamefont{N.}~\bibnamefont{de~Leon~Snapp}},
  \bibinfo{author}{\bibfnamefont{A.~V.} \bibnamefont{Akimov}},
  \bibinfo{author}{\bibfnamefont{M.-H.} \bibnamefont{Jo}},
  \bibinfo{author}{\bibfnamefont{M.~D.} \bibnamefont{Lukin}}, \bibnamefont{and}
  \bibinfo{author}{\bibfnamefont{H.}~\bibnamefont{Park}},
  \bibinfo{journal}{Nature Physics} \textbf{\bibinfo{volume}{5}},
  \bibinfo{pages}{475} (\bibinfo{year}{2009}).

\bibitem[{\citenamefont{Dufaux et~al.}(2010)\citenamefont{Dufaux, Dorfm\"uller,
  Vogelgesang, Burghard, and Kern}}]{DDV10}
\bibinfo{author}{\bibfnamefont{T.}~\bibnamefont{Dufaux}},
  \bibinfo{author}{\bibfnamefont{J.}~\bibnamefont{Dorfm\"uller}},
  \bibinfo{author}{\bibfnamefont{R.}~\bibnamefont{Vogelgesang}},
  \bibinfo{author}{\bibfnamefont{M.}~\bibnamefont{Burghard}}, \bibnamefont{and}
  \bibinfo{author}{\bibfnamefont{K.}~\bibnamefont{Kern}},
  \bibinfo{journal}{Appl.\ Phys.\ Lett.} \textbf{\bibinfo{volume}{97}},
  \bibinfo{pages}{161110} (\bibinfo{year}{2010}).

\bibitem[{\citenamefont{Goykhman et~al.}(2011)\citenamefont{Goykhman, Desiatov,
  Khurgin, Shappir, and Levy}}]{GDK11}
\bibinfo{author}{\bibfnamefont{I.}~\bibnamefont{Goykhman}},
  \bibinfo{author}{\bibfnamefont{B.}~\bibnamefont{Desiatov}},
  \bibinfo{author}{\bibfnamefont{J.}~\bibnamefont{Khurgin}},
  \bibinfo{author}{\bibfnamefont{J.}~\bibnamefont{Shappir}}, \bibnamefont{and}
  \bibinfo{author}{\bibfnamefont{U.}~\bibnamefont{Levy}},
  \bibinfo{journal}{Nano\ Lett.} \textbf{\bibinfo{volume}{11}},
  \bibinfo{pages}{2219} (\bibinfo{year}{2011}).

\bibitem[{\citenamefont{Knight et~al.}(2011)\citenamefont{Knight, Sobhani,
  Nordlander, and Halas}}]{KSN11}
\bibinfo{author}{\bibfnamefont{M.~W.} \bibnamefont{Knight}},
  \bibinfo{author}{\bibfnamefont{H.}~\bibnamefont{Sobhani}},
  \bibinfo{author}{\bibfnamefont{P.}~\bibnamefont{Nordlander}},
  \bibnamefont{and} \bibinfo{author}{\bibfnamefont{N.~J.} \bibnamefont{Halas}},
  \bibinfo{journal}{Science} \textbf{\bibinfo{volume}{332}},
  \bibinfo{pages}{702} (\bibinfo{year}{2011}).

\bibitem[{\citenamefont{Goodfellow et~al.}(2015)\citenamefont{Goodfellow,
  Chakraborty, Beams, Novotny, and Vamivakas}}]{GCB15}
\bibinfo{author}{\bibfnamefont{K.~M.} \bibnamefont{Goodfellow}},
  \bibinfo{author}{\bibfnamefont{C.}~\bibnamefont{Chakraborty}},
  \bibinfo{author}{\bibfnamefont{R.}~\bibnamefont{Beams}},
  \bibinfo{author}{\bibfnamefont{L.}~\bibnamefont{Novotny}}, \bibnamefont{and}
  \bibinfo{author}{\bibfnamefont{A.~N.} \bibnamefont{Vamivakas}},
  \bibinfo{journal}{Nano\ Lett.} \textbf{\bibinfo{volume}{15}},
  \bibinfo{pages}{5477} (\bibinfo{year}{2015}).

\bibitem[{\citenamefont{Britnell et~al.}(2013)\citenamefont{Britnell,
  Gorbachev, Geim, Ponomarenko, Mishchenko, Greenaway, Fromhold, Novoselov, and
  Eaves}}]{BGG13}
\bibinfo{author}{\bibfnamefont{L.}~\bibnamefont{Britnell}},
  \bibinfo{author}{\bibfnamefont{R.~V.} \bibnamefont{Gorbachev}},
  \bibinfo{author}{\bibfnamefont{A.~K.} \bibnamefont{Geim}},
  \bibinfo{author}{\bibfnamefont{L.~A.} \bibnamefont{Ponomarenko}},
  \bibinfo{author}{\bibfnamefont{A.}~\bibnamefont{Mishchenko}},
  \bibinfo{author}{\bibfnamefont{M.~T.} \bibnamefont{Greenaway}},
  \bibinfo{author}{\bibfnamefont{T.~M.} \bibnamefont{Fromhold}},
  \bibinfo{author}{\bibfnamefont{K.~S.} \bibnamefont{Novoselov}},
  \bibnamefont{and} \bibinfo{author}{\bibfnamefont{L.}~\bibnamefont{Eaves}},
  \bibinfo{journal}{Nat.\ Commun.} \textbf{\bibinfo{volume}{4}},
  \bibinfo{pages}{1794} (\bibinfo{year}{2013}).

\bibitem[{\citenamefont{Jeong et~al.}(2016)\citenamefont{Jeong, Oh, Bang,
  Jeong, An, Han, Kim, Yun, Kim, Park et~al.}}]{JOB16}
\bibinfo{author}{\bibfnamefont{H.}~\bibnamefont{Jeong}},
  \bibinfo{author}{\bibfnamefont{H.~M.} \bibnamefont{Oh}},
  \bibinfo{author}{\bibfnamefont{S.}~\bibnamefont{Bang}},
  \bibinfo{author}{\bibfnamefont{H.~J.} \bibnamefont{Jeong}},
  \bibinfo{author}{\bibfnamefont{S.-J.} \bibnamefont{An}},
  \bibinfo{author}{\bibfnamefont{G.~H.} \bibnamefont{Han}},
  \bibinfo{author}{\bibfnamefont{H.}~\bibnamefont{Kim}},
  \bibinfo{author}{\bibfnamefont{S.~J.} \bibnamefont{Yun}},
  \bibinfo{author}{\bibfnamefont{K.~K.} \bibnamefont{Kim}},
  \bibinfo{author}{\bibfnamefont{J.~C.} \bibnamefont{Park}},
  \bibnamefont{et~al.}, \bibinfo{journal}{Nano\ Lett.}
  \textbf{\bibinfo{volume}{16}}, \bibinfo{pages}{1858} (\bibinfo{year}{2016}).

\bibitem[{\citenamefont{Katkov and Osipov}(2017)}]{KO17}
\bibinfo{author}{\bibfnamefont{V.~L.} \bibnamefont{Katkov}} \bibnamefont{and}
  \bibinfo{author}{\bibfnamefont{V.~A.} \bibnamefont{Osipov}},
  \bibinfo{journal}{J.\ Vac.\ Sci.\ Technol.\ B} \textbf{\bibinfo{volume}{35}},
  \bibinfo{pages}{050801} (\bibinfo{year}{2017}).

\bibitem[{\citenamefont{{Castro Neto} et~al.}(2009)\citenamefont{{Castro Neto},
  Guinea, Peres, Novoselov, and Geim}}]{CGP09}
\bibinfo{author}{\bibfnamefont{A.~H.} \bibnamefont{{Castro Neto}}},
  \bibinfo{author}{\bibfnamefont{F.}~\bibnamefont{Guinea}},
  \bibinfo{author}{\bibfnamefont{N.~M.~R.} \bibnamefont{Peres}},
  \bibinfo{author}{\bibfnamefont{K.~S.} \bibnamefont{Novoselov}},
  \bibnamefont{and} \bibinfo{author}{\bibfnamefont{A.~K.} \bibnamefont{Geim}},
  \bibinfo{journal}{Rev.\ Mod.\ Phys.} \textbf{\bibinfo{volume}{81}},
  \bibinfo{pages}{109} (\bibinfo{year}{2009}).

\bibitem[{\citenamefont{Wallace}(1947)}]{W1947}
\bibinfo{author}{\bibfnamefont{P.~R.} \bibnamefont{Wallace}},
  \bibinfo{journal}{Phys.\ Rev.} \textbf{\bibinfo{volume}{71}},
  \bibinfo{pages}{622} (\bibinfo{year}{1947}).

\bibitem[{\citenamefont{Wunsch et~al.}(2006)\citenamefont{Wunsch, Stauber,
  Sols, and Guinea}}]{WSS06}
\bibinfo{author}{\bibfnamefont{B.}~\bibnamefont{Wunsch}},
  \bibinfo{author}{\bibfnamefont{T.}~\bibnamefont{Stauber}},
  \bibinfo{author}{\bibfnamefont{F.}~\bibnamefont{Sols}}, \bibnamefont{and}
  \bibinfo{author}{\bibfnamefont{F.}~\bibnamefont{Guinea}},
  \bibinfo{journal}{New\ J.\ Phys.} \textbf{\bibinfo{volume}{8}},
  \bibinfo{pages}{318} (\bibinfo{year}{2006}).

\bibitem[{\citenamefont{Hwang and {Das Sarma}}(2007)}]{HD07}
\bibinfo{author}{\bibfnamefont{E.~H.} \bibnamefont{Hwang}} \bibnamefont{and}
  \bibinfo{author}{\bibfnamefont{S.}~\bibnamefont{{Das Sarma}}},
  \bibinfo{journal}{Phys.\ Rev.\ B} \textbf{\bibinfo{volume}{75}},
  \bibinfo{pages}{205418} (\bibinfo{year}{2007}).

\bibitem[{\citenamefont{Pettit et~al.}(1975)\citenamefont{Pettit, Silcox, and
  Vincent}}]{PSV1975}
\bibinfo{author}{\bibfnamefont{R.~B.} \bibnamefont{Pettit}},
  \bibinfo{author}{\bibfnamefont{J.}~\bibnamefont{Silcox}}, \bibnamefont{and}
  \bibinfo{author}{\bibfnamefont{R.}~\bibnamefont{Vincent}},
  \bibinfo{journal}{Phys.\ Rev.\ B} \textbf{\bibinfo{volume}{11}},
  \bibinfo{pages}{3116} (\bibinfo{year}{1975}).

\bibitem[{\citenamefont{Ihn et~al.}(2010)\citenamefont{Ihn, G\"uttinger,
  Molitor, Schnez, Schurtenberger, Jacobsen, Hellm\"uller, Frey, Dr\"oscher,
  Stampfer et~al.}}]{IGM10}
\bibinfo{author}{\bibfnamefont{T.}~\bibnamefont{Ihn}},
  \bibinfo{author}{\bibfnamefont{J.}~\bibnamefont{G\"uttinger}},
  \bibinfo{author}{\bibfnamefont{F.}~\bibnamefont{Molitor}},
  \bibinfo{author}{\bibfnamefont{S.}~\bibnamefont{Schnez}},
  \bibinfo{author}{\bibfnamefont{E.}~\bibnamefont{Schurtenberger}},
  \bibinfo{author}{\bibfnamefont{A.}~\bibnamefont{Jacobsen}},
  \bibinfo{author}{\bibfnamefont{S.}~\bibnamefont{Hellm\"uller}},
  \bibinfo{author}{\bibfnamefont{T.}~\bibnamefont{Frey}},
  \bibinfo{author}{\bibfnamefont{S.}~\bibnamefont{Dr\"oscher}},
  \bibinfo{author}{\bibfnamefont{C.}~\bibnamefont{Stampfer}},
  \bibnamefont{et~al.}, \bibinfo{journal}{Mater.\ Today}
  \textbf{\bibinfo{volume}{13}}, \bibinfo{pages}{44} (\bibinfo{year}{2010}).

\bibitem[{\citenamefont{Low et~al.}(2014)\citenamefont{Low, Rold\'an, Wang,
  Xia, Avouris, {Mart\'{\i}n Moreno}, and Guinea}}]{LRW14}
\bibinfo{author}{\bibfnamefont{T.}~\bibnamefont{Low}},
  \bibinfo{author}{\bibfnamefont{R.}~\bibnamefont{Rold\'an}},
  \bibinfo{author}{\bibfnamefont{H.}~\bibnamefont{Wang}},
  \bibinfo{author}{\bibfnamefont{F.}~\bibnamefont{Xia}},
  \bibinfo{author}{\bibfnamefont{P.}~\bibnamefont{Avouris}},
  \bibinfo{author}{\bibfnamefont{L.}~\bibnamefont{{Mart\'{\i}n Moreno}}},
  \bibnamefont{and} \bibinfo{author}{\bibfnamefont{F.}~\bibnamefont{Guinea}},
  \bibinfo{journal}{Phys.\ Rev.\ Lett.} \textbf{\bibinfo{volume}{113}},
  \bibinfo{pages}{106802} (\bibinfo{year}{2014}).

\bibitem[{\citenamefont{Huber et~al.}(2017)\citenamefont{Huber, Mooshammer,
  Plankl, Viti, Sandner, Kastner, Frank, Fabian, Vitiello, Cocker
  et~al.}}]{HMP17}
\bibinfo{author}{\bibfnamefont{M.~A.} \bibnamefont{Huber}},
  \bibinfo{author}{\bibfnamefont{F.}~\bibnamefont{Mooshammer}},
  \bibinfo{author}{\bibfnamefont{M.}~\bibnamefont{Plankl}},
  \bibinfo{author}{\bibfnamefont{L.}~\bibnamefont{Viti}},
  \bibinfo{author}{\bibfnamefont{F.}~\bibnamefont{Sandner}},
  \bibinfo{author}{\bibfnamefont{L.~Z.} \bibnamefont{Kastner}},
  \bibinfo{author}{\bibfnamefont{T.}~\bibnamefont{Frank}},
  \bibinfo{author}{\bibfnamefont{J.}~\bibnamefont{Fabian}},
  \bibinfo{author}{\bibfnamefont{M.~S.} \bibnamefont{Vitiello}},
  \bibinfo{author}{\bibfnamefont{T.~L.} \bibnamefont{Cocker}},
  \bibnamefont{et~al.}, \bibinfo{journal}{Nat.\ Nanotech.}
  \textbf{\bibinfo{volume}{12}}, \bibinfo{pages}{207} (\bibinfo{year}{2017}).

\bibitem[{\citenamefont{Liu et~al.}(2013)\citenamefont{Liu, Shan, Yao, Yao, and
  Xiao}}]{LSY13}
\bibinfo{author}{\bibfnamefont{G.-B.} \bibnamefont{Liu}},
  \bibinfo{author}{\bibfnamefont{W.-Y.} \bibnamefont{Shan}},
  \bibinfo{author}{\bibfnamefont{Y.}~\bibnamefont{Yao}},
  \bibinfo{author}{\bibfnamefont{W.}~\bibnamefont{Yao}}, \bibnamefont{and}
  \bibinfo{author}{\bibfnamefont{D.}~\bibnamefont{Xiao}},
  \bibinfo{journal}{Phys.\ Rev.\ B} \textbf{\bibinfo{volume}{88}},
  \bibinfo{pages}{085433} (\bibinfo{year}{2013}).

\bibitem[{\citenamefont{Yu et~al.}(2009)\citenamefont{Yu, Zhao, Ryu, Brus, Kim,
  and Kim}}]{YZR09}
\bibinfo{author}{\bibfnamefont{Y.-J.} \bibnamefont{Yu}},
  \bibinfo{author}{\bibfnamefont{Y.}~\bibnamefont{Zhao}},
  \bibinfo{author}{\bibfnamefont{S.}~\bibnamefont{Ryu}},
  \bibinfo{author}{\bibfnamefont{L.~E.} \bibnamefont{Brus}},
  \bibinfo{author}{\bibfnamefont{K.~S.} \bibnamefont{Kim}}, \bibnamefont{and}
  \bibinfo{author}{\bibfnamefont{P.}~\bibnamefont{Kim}},
  \bibinfo{journal}{Nano\ Lett.} \textbf{\bibinfo{volume}{9}},
  \bibinfo{pages}{3430} (\bibinfo{year}{2009}).

\bibitem[{\citenamefont{Clementi and Roetti}(1974)}]{CR1974}
\bibinfo{author}{\bibfnamefont{E.}~\bibnamefont{Clementi}} \bibnamefont{and}
  \bibinfo{author}{\bibfnamefont{C.}~\bibnamefont{Roetti}},
  \bibinfo{journal}{At.\ Data\ Nucl.\ Data\ Tables}
  \textbf{\bibinfo{volume}{14}}, \bibinfo{pages}{177} (\bibinfo{year}{1974}).

\bibitem[{\citenamefont{Watanabe et~al.}(2004)\citenamefont{Watanabe,
  Taniguchi, and Kanda}}]{WTK04}
\bibinfo{author}{\bibfnamefont{K.}~\bibnamefont{Watanabe}},
  \bibinfo{author}{\bibfnamefont{T.}~\bibnamefont{Taniguchi}},
  \bibnamefont{and} \bibinfo{author}{\bibfnamefont{H.}~\bibnamefont{Kanda}},
  \bibinfo{journal}{Nat.\ Mater.} \textbf{\bibinfo{volume}{3}},
  \bibinfo{pages}{404} (\bibinfo{year}{2004}).

\bibitem[{\citenamefont{Geick et~al.}(1966)\citenamefont{Geick, Perry, and
  Rupprecht}}]{GPR1960}
\bibinfo{author}{\bibfnamefont{R.}~\bibnamefont{Geick}},
  \bibinfo{author}{\bibfnamefont{C.~H.} \bibnamefont{Perry}}, \bibnamefont{and}
  \bibinfo{author}{\bibfnamefont{G.}~\bibnamefont{Rupprecht}},
  \bibinfo{journal}{Phys.\ Rev.} \textbf{\bibinfo{volume}{146}},
  \bibinfo{pages}{543} (\bibinfo{year}{1966}).

\end{thebibliography}

\end{document}